\magnification=\magstephalf
\input amstex
\loadbold
\documentstyle{amsppt}
\refstyle{A}
\NoBlackBoxes

\vsize=7.5in

\def\pf{\hfill $\square$}
\def\c{\cite}

\def\fg{\frak{g}}
\def\fh{\frak{h}}

\def\end{\text{End}}
\def\group{\{\cdot , \cdot\}}
\def\bi{\boldkey i}
\def\bp{\boldkey Pr}
\def\bm{\boldkey m}
\def\bL{\boldkey L}
\def\bpr{\boldkey pr}
\def\varphit{\widetilde \varphi}
\def\psit{\widetilde \psi}

\def\TF_1{\widetilde F_{1}}
\def\TF_2{\widetilde F_{2}}
\def\TF{\widetilde F}

\topmatter
\title Poisson involutions, Spin Calogero-Moser systems associated
with symmetric Lie subalgebras and the symmetric space   
spin Ruijsenaars-Schneider models  \endtitle
\leftheadtext{L.-C. Li}
\rightheadtext{Poisson involutions, spin CM and RS}

\author Luen-Chau Li\endauthor
\address{L.-C. Li, Department of Mathematics,Pennsylvania State University
University Park, PA  16802, USA}\endaddress
\email luenli\@math.psu.edu\endemail
\abstract  We develop a general scheme to 
construct integrable systems starting from realizations in 
symmetric coboundary dynamical Lie algebroids and symmetric
coboundary Poisson groupoids.  The method is based on the successive 
use of Dirac reduction and Poisson reduction.  Then we show that
certain spin Calogero-Moser systems associated with symmetric
Lie subalgebras can be studied in this fashion.  We also consider
some spin-generalized Ruisjenaars-Schneider equations which correspond
to the $N$-soliton solutions of $A^{(1)}_n$ affine Toda field theory.
In this case, we show how the equations are obtained from the Dirac 
reduction of some Hamiltonian 
system on a symmetric coboundary dynamical Poisson groupoid.
\endabstract
\endtopmatter

\document
\subhead
1. \ Introduction
\endsubhead

\baselineskip 15pt
\bigskip

In the last few years, a groupoid-theoretic scheme based on
the coboundary dynamical Poisson groupoids and their corresponding
Lie bialgebroids was introduced in the study of certain integrable
Hamiltonian systems and their solutions \c{LX},\c{L1},\c{L2},\c{L3}.
As is well-known, these geometric objects are naturally associated
with so-called classical dynamical r-matrices \c{EV},\c{BKS} which
first appeared in the context of Wess-Zumino-Witten (WZW) conformal
field theory \c{BDF},\c{F}.

In this paper, we shall continue to use these geometric objects
to unify the study of a variety of Hamiltonian systems
known under the general name of spin Calogero-Moser systems
and spin Ruijsenaars-Schneider models.  More specifically, 
we shall consider in this work examples of such systems which turn out
to be realizable in the stable loci of the geometric objects mentioned
above under Poisson involutions.  As we know through the work
in \c{X}, the stable locus of a Poisson involution is an example of 
a class of submanifolds with induced Poisson structures which the author in 
\c{X} called Dirac submanifolds.(The fact that the stable locus of a
Poisson involution carries a natural induced Poisson structure was
also noted by the authors in \c{FV}.)  Indeed, as we shall explain
below, it is advantageous to formulate several of our results in 
this broader framework.

We now give an outline of our approach. As starting point, we consider
certain spin Calogero-Moser
systems (resp.~spin Ruijsenaars-Schneider models) which can be
realized in the dual bundles of symmetric coboundary dynamical Lie algebroids
(resp.~ symmetric coboundary dynamical Poisson groupoids).  
For these Hamiltonian systems, the underlying Poisson
manifolds as well as their realization spaces are both Hamiltonian
$H$-spaces which carry natural Poisson involutions. The construction of 
the integrable systems of
interests then proceeds in two stages.  In the first stage, we apply 
Dirac reduction (which will be developed here) to 
reduce the initial realization maps to ones between the stable loci 
of Poisson involutions.  In this way, we obtain the realization of
the Dirac reduction of the afore-mentioned systems.  In general, these 
reduced systems are not integrable systems (see Theorem 3.15,
Theorem 3.18 and Section 4.2 for exceptions).  However, as it turns out, the 
stable loci are Hamiltonian $D$-spaces for some subgroup $D$ of $H$.
Moreover, the natural invariant functions Poisson commute on certain 
fibers of the equivariant momentum maps.  Consequently, we can apply 
Poisson reduction (and this is the second stage) to obtain the associated 
integrable systems.

There are several motivations for this work.  One of these has come
from the desire to understand the Hamiltonian formulation as
well as the integrability of the equations of motion which arise
from the so-called level dynamics approach in random matrix theory
\c{Y},\c{HKS},\c{NM},\c{GRMN}. This connection is reflected in our
choice of examples in Section 4.  On the other hand, there is
a well-known correspondence between the $N$-soliton solutions of the
$A^{(1)}_n$ affine Toda field theory and some spin-generalized 
Ruijensaars-Schneider equations \c{BH}. (Some of the variables
in these equations actually depend on the choice of eigenvectors of
a certain skew-Hermitian matrix $V$.) However, the Hamiltonian
formulation of these equations has remained open.  We shall give
a solution to this problem in Section 5 below.  As the reader will
see, these equations are related to a symmetric coboundary dynamical 
Poisson groupoid $(\Gamma,\Sigma)$, where 
$\Gamma$ is associated to a hyperbolic dynamical r-matrix, and
$\Sigma$ is a Poisson involution on $\Gamma.$  More precisely, they 
can be obtained from a 
Hamiltonian system on the stable locus $\Gamma^{\Sigma}$ of 
$\Sigma$ by restricting the equations of motion to an appropriate
fiber of the momentum map.  Consequently, the system which is
invariant under the gauge freedom in picking the eigenvectors
of $V$ is an integrable Hamiltonian system
on a Poisson reduction of $\Gamma^{\Sigma}.$  Finally we remark that
in the process of assembling the necessary machinery in order to
tackle the above problems, we will give the explicit expression
for the Poisson structure on the stable locus of a Poisson involution
on a coboundary dynamical Lie algebroid (resp.~coboundary dynamical
Poisson groupoid).  Thus this answers a question raised in
\c{X}.
  
The paper is organized as follows.  In section 2, we begin by recalling some
basic facts about coboundary dynamical Lie algebroids and coboundary
dynamical Poisson groupoids which will be used throughout the paper.
In particular, we will discuss a subclass of such Lie algebroids defined by 
so-called classical dynamical r-matrices with spectral parameter. We will
also recall what we mean by spin Calogero-Moser systems associated
with this subclass of coboundary dynamical Lie algebroids. In section 3, the 
main goal is to develop a general 
scheme of constructing integrable systems based on realization in
symmetric coboundary dynamical Poisson groupoids and the dual bundles
of symmetric coboundary dynamical Lie algebroids.  As we already mentioned 
above in the context of specific examples, the construction proceeds 
in two stages.  For Dirac reduction, our main tool comes from an
elementary result which shows how to reduce a Poisson map 
between two Poisson manifolds to one between their respective Dirac
submanifolds (Theorem 3.2 and Corollary 3.5). From this, we also
obtain a condition under which a Dirac submanifold $Q$ of a 
Hamiltonian $G$-space $P$ is Hamiltonian $H$-space for some 
Lie subgroup $H$ of $G$ (Proposition 3.6 and Corollary 3.7).
There are two reasons for formulating our results in terms
of Dirac submanifolds.  First, the notion offers a better conceptual
framework.  Secondly, when formulated in this broader framework, 
the results are also applicable to the cosymplectic submanifolds \c{W1}
(when $P$ is symplectic, the cosymplectic submanifolds of $P$
are precisely its symplectic submanifolds).
In the special case when the Dirac submanifold is given by the stable locus 
of a Poisson involution on the dual bundle of a coboundary dynamical Lie 
algebroid (resp.~coboundary 
dynamical Poisson groupoid), we also derive the intrinsic expression for 
the induced Poisson structure which is essential for our purpose here.
In Section 4, we introduce several examples of spin Calogero-Moser
systems associated with real symmetric Lie algebras.  Then we show how
the reduction procedure developed in Section 3 can be carried out
to obtain the associated integrable systems of interests.  In the special 
case when the Lie
algebra $\fg$ is $gl(N,\Bbb{C})$, we also provide a sketch of the Liouville
integrability of the associated integrable models.  Note that 
our  goal of this section is suggestive
rather than exhaustive in the sense that we have  made no attempt
to give a classification of systems which can be treated by our
method.  Finally, in Section 5, we consider
the spin Ruijsenaars-Schneider models associated with a symmetric
coboundary dynamical Poisson groupoid $(\Gamma,\Sigma)$.  In 
this case, the realization map is just the identity map and
it is easy to show how the scheme in Section 3 can be implemented.
As we mentioned earlier, our goal here is explain how the 
spin-generalized Ruijsenaars-Schneider equations in
\c{BH} are obtained from an invariant  Hamiltonian system on 
$\Gamma^{\Sigma}$ which is
a special case of what we call symmetric space Ruijsenaars-Schneider
models here.

To close, we remark that a factorization theory also exists for the solution
of the Hamiltonian systems treated here (provided the classical
dynamical r-matrix satisfies the modified dynamical Yang-Baxter
equation), as is clear from assumptions
A5 and G5 in Section 3 and the development in \c{L1},\c{L2}.  For this
reason, we do not give any details here.
\bigskip
\noindent{\bf Acknowledgments.} The author would like to thank Ping Xu for the
reference \c{FV} when this work was in its final stage of preparation.
\bigskip
\bigskip

\subhead
2. \ Preliminaries
\endsubhead
\bigskip

The purpose of this section is to recall some basic results about
coboundary dynamical Lie algebroids and coboundary dynamical Poisson
groupoids.  For our applications in this work, we will pay special attention 
to a subclass of such Lie algebroids which are associated with so-called
classical dynamical r-matrices with spectral parameter.  We will also recall
what we mean by spin Calogero-Moser systems associated with this subclass of
coboundary dynamical Lie algebroids.

Let $G$ be a connected Lie group, and $H\subset G$ a connected Lie
subgroup.  We shall denote by $\fg$ and $\fh$ the corresponding Lie
algebras and let $\iota :\fh \longrightarrow \fg$ be the Lie inclusion.
In what follows, the Lie groups and Lie algebras can be real or
complex unless we specify otherwise.

We begin by recalling a fundamental construction in \c{EV} which gives
a geometric interpretation of dynamical r-matrices in terms of Poisson
groupoids. We shall, however, follow the formulation in \c{L1} and in 
particular we shall give the explicit expression for the Poisson
structure which is essential for our purpose here.
Let $U\subset \fh^*$ be a connected 
$Ad_H^*$-invariant open subset, we say that a smooth (resp.~holomorphic)
map 
$R:U\longrightarrow L(\fg^*, \fg)$  (here and henceforth 
we denote by $L(\fg^*,\fg)$ the set of linear maps from $\fg^*$ to $\fg$)
is a classical dynamical r-matrix if and only if it is pointwise
skew-symmetric:
$$<R(q)(A), B>=- <A, R (q) B>\eqno (2.1)$$
and satisfies the classical dynamical Yang-Baxter condition
$$\eqalign {&[R(q)A, R(q)B] +R(q)(ad^*_{R(q)A}B-ad^*_{R(q)B}A)\cr
+&dR(q)\iota^*A(B) - dR(q)\iota^*B(A) + <dR(q)(\cdot)(A),B> 
= \chi (A,B),\cr} \eqno (2.2)$$
where $ <dR(q)(\cdot)(A),B>$ is the element in $\fh$ whose
pairing with $\lambda\in \fh^*$ is given by
$<dR(q)(\lambda)(A),B>$ and
$\chi : \fg^*\times \fg^* \longrightarrow  \fg$ is
$G$-equivariant, that is,
$$ \chi (Ad^*_{g^{-1}}A, Ad^*_{g^{-1}}B) = Ad_g\, \chi(A,B)\eqno(2.3)$$
for all $A, B\in \fg^*$, $g\in G$, and all $q\in U$.
\smallskip
The dynamical $r$-matrix is said to be $H$-equivariant
if and only if
$$R(Ad^*_{h^{-1}} q)=Ad_h \circ R(q)\circ Ad^*_h \eqno (2.4)$$
for all $h\in H, q\in U$.

We shall equip $\Gamma = U\times G\times U$ with the trivial Lie 
groupoid structure over $U$ \c{M} with target and source maps
$$\alpha (u,g,v)= u, \quad  \beta(u,g,v)= v \eqno(2.5)$$
and multiplication map
$$m((u,g,v), (v,g',w))= (u,gg',w).\eqno(2.6)$$

For a smooth (resp.~holomorphic) function $\varphi$ on $\Gamma$, we define 
its partial derivatives
and its left and right gradients (with respect to $G$) by 
$$\eqalign {&<\delta_1 \varphi, u'>= {d\over dt}{\Big|_{t=0}} 
\varphi(u+tu', g, v),\quad <\delta_2 \varphi, v'>= {d\over dt}{\Big|_{t=0}} 
\varphi(u, g, v+tv'),\,
u', v'\in \fh^{*}\cr
&<D\varphi, X>= {d\over dt}{\Big|_{t=0}} \varphi(u, e^{tX}g,v),
\quad <D' \varphi, X>= {d\over dt}{\Big|_{t=0}} 
\varphi(u, ge^{tX}, v), \, X\in \fg.\cr}$$

\proclaim
{Theorem 2.1} (a) The  bracket 
$$\eqalign {\{\varphi,\psi\}_R (u,g,v)=& <u, [\delta_1\varphi,\,
\delta_1\psi]> -<v, [\delta_2\varphi,\, \delta_2\psi]>\cr
& - <\iota\delta_1 \varphi,\, D\psi> - <\iota\delta_2 \varphi, D'\psi>\cr
& + <\iota\delta_1 \psi,\, D\varphi> + <\iota\delta_2 \psi , D'\varphi>\cr
& +<R(v) D'\varphi, D'\psi> - <R(u) D\varphi, D\psi>\cr}\eqno(2.7)
$$
defines a Poisson structure on $\Gamma$ if and only if
$R: U\longrightarrow L(\fg^*, \fg)$ is an $H$-
equivariant classical dynamical $r$-matrix.
\smallskip
\noindent (b) The trivial Lie groupoid $\Gamma$ equipped with the Poisson 
bracket
$\{\, \cdot, \cdot \,\}_R$ is a Poisson groupoid.  Moreover, it is
a Hamiltonian $H$-space under the natural left and right $H$-actions 
with equivariant momentum maps given by $\alpha$ and $\beta$ respectively.
\endproclaim
\smallskip
We shall call the pair $(\Gamma, \{\,\cdot, \cdot \,\}_R)$ the {\it coboundary 
dynamical Poisson groupoid \/} associated to $R$.  Note that the 
explicit expression for $\{\varphi,\psi\}_R$ in Theorem 2.1(a) above
can be derived from
the characterizing properties in \c{EV} and the corresponding
expression for the general dynamical case can be found in \c{LP}.

Let $A\Gamma = \bigcup_{q\in U}\lbrace 0_{q}\rbrace\times \fg\times \fh^*\simeq
TU\times \fg$ be the Lie algebroid of the trivial Lie groupoid $\Gamma$.
(See \c{CdSW},\c{M} for details.)  Then associated with 
 $(\Gamma, \{\,\cdot, \cdot \,\}_R)$ is 
a Lie algebroid structure on the dual bundle 
$A^*\Gamma = \bigcup_{q\in U} \lbrace 0_{q} \rbrace \times \fg^* \times 
\fh \simeq T^{*}U\times \fg^*$ as a consequence of Weinstein's 
coisotropic calculus \c{W2}.(See \c{LP} for a more general discussion
and \c{BKS} for a different approach.)  The anchor map 
$a_{*}:A^{*}\Gamma\longrightarrow TU$ of $A^*\Gamma$ of this Lie 
algebroid is given by
$$a_{*} (0_q,A,Z)=(q, \iota^{*}A-ad^{*}_{Z}q)\eqno(2.8)$$
while the bracket $[\cdot, \cdot]_{A^*\Gamma}$ on 
$Sect(U,A^{*}\Gamma)$ has the following form 
\c{BKS},\c{L2}:
$$\eqalign{& [(0,\xi,Z),(0,\xi',Z')]_{A^{*}\Gamma} (q)\cr
     = & (0_q, d\xi'(q)(\iota^{*}\xi(q)-ad^{*}_{Z(q)} q)
         -d\xi(q)(\iota^{*} \xi'(q)-ad^{*}_{Z'(q)} q)\cr
       &\quad -ad^{*}_{R(q)\xi(q)-Z(q)} \xi'(q)
         +ad^{*}_{R(q)\xi'(q)-Z'(q)} \xi(q),\cr
       &\quad \,dZ'(q)(\iota^{*}\xi(q)-ad^{*}_{Z(q)} q)
         -dZ(q)(\iota^{*} \xi'(q)-ad^{*}_{Z'(q)} q)\cr
       &\quad -[Z,Z'](q) + <dR(q)(\cdot)\xi(q),\xi'(q)>)\cr}\eqno(2.9)
$$
where $\xi,\xi':U\longrightarrow \fg^{*}$, $Z,Z':U\longrightarrow \fh$ are
smooth (resp.~holomorphic) maps and $<dR(q)(\cdot)\xi(q),\xi'(q)>$ 
is the element in $\fh$
whose pairing with $\lambda \in \fh^*$ is $<dR(q)(\lambda)\xi(q),\xi'(q)>.$
We shall call $(A^*\Gamma, [\cdot,\cdot]_{A^*\Gamma}, a_{*})$ the
{\it coboundary dynamical Lie algebroid\/} associated to $R.$

Now, for any Lie algebroid $(A,[\cdot,\cdot]_{A}, a_A)$ over
a smooth manifold $M$, recall that there exists a Lie-Poisson structure 
on the dual bundle $A^*$ \c{CDW} which is uniquely
determined by the property 
$$\{l_{X},l_{Y}\} = l_{[X,Y]_{A}} \eqno(2.10)$$
where for $X,Y\in Sect(M,A)$, $l_{X}$ and $l_{Y}$ are the corresponding
linear functions on $A^*$.  The following result was obtained in
\c{L2}.

\proclaim
{Theorem 2.2} (a) The Lie-Poisson structure on the dual bundle $A\Gamma$
of the coboundary dynamical Lie algebroid
$(A^{*}\Gamma, [\cdot,\cdot]_{A^{*}\Gamma},a_{*})$
is given by 
$$\eqalign{& \{\varphi, \psi\}_{A\Gamma} (q,\lambda,X)\cr
  =\,&-<\lambda, [\delta_{2}\varphi,\delta_{2}\psi]>+
     <dR(q)(\lambda)\delta \varphi,\delta\psi>\cr
     & +<X, -ad^{*}_{R(q)\delta \varphi-\delta_{2}\varphi}
          \delta \psi + ad^{*}_{R(q)\delta \psi-\delta_{2}\psi}
          \delta \varphi >\cr
     &-<q,[\delta_{2}\varphi,\delta_{1}\psi]+[\delta_{1}\varphi,
       \delta_{2}\psi]>+<\delta_{1}\psi, \iota^{*}\delta \varphi>
       -<\delta_{1}\varphi,\iota^{*}\delta\psi>.\cr}\eqno(2.11)$$
\smallskip
\noindent (b)  With the action of $H$ on $A\Gamma$ defined by the
formula 
$$h\cdot(q,\lambda,X)=(Ad^{*}_{h^{-1}}q, Ad^{*}_{h^{-1}}\lambda,Ad_{h}X),
\eqno(2.12)$$ the
dual bundle $A\Gamma$ of the coboundary dynamical Lie algebroid
$A^{*}\Gamma$ equipped with the Lie-Poisson structure is a Hamiltonian 
$H$-space with equivariant momentum
map 
$$\gamma:A\Gamma\longrightarrow \fh^{*}, \,\,(q,\lambda,X)\mapsto
\lambda.\eqno(2.13)$$
\endproclaim

In the rest of the section, we shall assume that 
$\fg$ is a Lie algebra with a
nondegenerate invariant pairing $(\cdot,\cdot)$ and $\fh\subset \fg$
is a non-degenerate (i.e. $(\cdot,\cdot)_{\fh\times \fh}$ is nondegenerate)
abelian Lie subalgebra.  Then we can make
the identifications $\fg^*\simeq \fg$, $\fh^*\simeq \fh$,
$ad^*\simeq -ad$, $\iota^*\simeq \Pi_{\fh}$, where $\Pi_{\fh}$ is
the projection map to $\fh$ relative to the direct sum
decomposition $\fg= \fh\oplus \fh^{\perp}$.  Hence we can regard
$R(q)$ as taking values in $End(\fg)$.  In this case, an important
sufficient condition for an $H$-equivariant map $R$ to define
a coboundary dynamical Poisson groupoid is given by the
modified dynamical Yang-Baxter equation (mDYBE):
$$\eqalign {& [R(q)X, R(q)Y]-R(q)([R(q)X,Y]+[X,R(q)Y])\cr
+\,&dR(q)\Pi_{\fh}X(Y) - dR(q)\Pi_{\fh}Y(X) + (dR(q)(\cdot)X,Y)\cr
=\,& -c^{2} [X,Y],\cr}\eqno(2.14)$$
where $c$ is a nonzero constant.  

We now turn our attention to
a subclass of $(\Gamma, \{\cdot,\cdot\}_{R})$ and
$(A^{*}\Gamma, [\cdot,\cdot]_{A^{*}\Gamma},a_{*})$
which are associated with so-called classical dynamical
r-matrices with spectral parameter.  In the following, we
shall assume that $\fg$ is a Lie algebra over $\Bbb{C}$.

\definition
{Definition 2.3 \c{EV},\c{LX}} A classical dynamical r-matrix with
spectral parameter is a meromorphic map 
$r:\fh \times \Bbb{C}\longrightarrow \fg\otimes\fg$ with
a simple pole at $z=0$ satisfying the following conditions
for all $(q,z)\in \fh\times \Bbb{C}$ away from the poles of $r$,
\smallskip
\noindent 1. the zero weight condition:
$$[h\otimes 1 + 1\otimes h, \,\,r(q,z)] = 0, \eqno(2.15)$$
for all $h\in \fh$,
\smallskip
\noindent 2. the generalized unitarity condition:
$$r^{12}(q,z) + r^{21}(q,-z) = 0,\eqno(2.16)$$
\smallskip
\noindent 3. the residue condition:
$$ Res_{z=0}\, r(q,z) = \Omega, \eqno(2.17)$$
where $\Omega\in (S^{2}\fg)^{\fg}$ is the Casimir element
corresponding to $(\cdot,\cdot)$,
\smallskip
\noindent 4. the classical dynamical Yang-Baxter equation (CDYBE) with
spectral parameter:
$$\eqalign{Alt(dr)&+\left[r^{12}(q,z_{12}),\,\, r^{13}(q,z_{13}) + r^{23}(q,z_{23})
  \right]\cr
  &+ \left[r^{13}(q,z_{13}),\, r^{23}(q,z_{23})\right] = 0,\cr}\eqno(2.19)$$
where $z_{ij}=z_i-z_j$.
\enddefinition

Let $L\fg$ be the loop algebra consisting of Laurent series with coefficients
in $\fg.$  If $r$ is a classical dynamical r-matrix with spectral parameter,
we define
$$(R(q)X)(z) = p.v. {1\over {2\pi i}}\oint_C (r(q,w-z), X(w)\otimes 1)dw, 
\, \, \, X\in L\fg,\eqno(2.20)$$
where $C$ is a small circle centered at $0$ with positive orientation, and
$p.v.$ denotes the principal value of the improper integral.
We have the following result \c{LX}.
\proclaim
{Theorem 2.4} (a) $R$ is an $H$-equivariant classical dynamical r-matrix
which satisfies the mDYBE with $c=-{1\over 4}.$
\smallskip
\noindent (b) For $X\in L\fg$, we have the formula
$$(R(q)X )(z)= {1\over 2} X(z) +
\sum_{k\geq 0}\frac{1}{k!}\left ( \frac{\partial^{k}r}{\partial z^{k}}
(q, -z ), \ X_{-(k+1)}\otimes 1 \right ).$$  
\endproclaim

We now fix an open connected set $U\subset \fh$ on which $R$ is 
holomorphic.  Let $A\Omega = U\times \fh\times \fg$ be the
trivial Lie algebroid over $U$ with vertex algebra $\fg$.
We shall identify its dual bundle $A^*\Omega$ with
$U\times\fh\times\fg$ and equip it with the Lie-Poisson
structure. On the other hand, we can use $R$ in Theorem 2.4
above to construct the associated coboundary dynamical Lie
algebroid $A^*\Gamma\simeq U\times \fh\times L\fg$.  Therefore,
we can equip its dual bundle 
$A\Gamma$ with the corresponding Lie-Poisson structure.
For each $q\in U$, we now define a map
$r^{\#}_{-}(q):\fg\longrightarrow L\fg$
by the formula
$$((r^{\#}_{-} (q)\xi)(z), \eta) = (r(q,z), \eta \otimes \xi) \eqno(2.21)$$
where $\xi$, $\eta \in \fg.$ 

\proclaim
{Theorem 2.5 \c{LX}} The map $\rho: A^*\Omega\longrightarrow A\Gamma$ given by
$$(q,p,\xi)\mapsto (q, -\Pi_{\fh}\xi, p + r^{\#}_{-}(q)\xi)$$
is an $H$-equivariant Poisson map, where $H$ acts on $A^*\Omega$
by $h\cdot(q,p,\xi) = (q,p,Ad_{h}\xi)$.
\endproclaim

\definition
{Definition 2.6} Let $r$ be a classical dynamical r-matrix with
spectral parameter and let $L= Pr_{3}\circ \rho$, where $\rho$
is the realization map in Theorem 2.5.  Then the (complex
holomorphic) Hamiltonian system on  $A^*\Omega$ generated by
the Hamiltonian function 
$${\Cal H}^{\Bbb{C}}(q,p,\xi) = {1\over 2}\oint_{C} (L(q,p,\xi),L(q,p,\xi))
  \frac{dz}{2\pi i z}\eqno(2.22)$$
is called the spin Calogero-Moser system associated with $r$. Here,
$C$ is a small circle centered at $0$ with the positive orientation.
\enddefinition

\smallskip
\noindent{\bf Remark 2.7.} Actually we will use the real version of
Theorem 2.5 in our application in Section 4 below.

\bigskip
\bigskip

\subhead
3. \ Dirac reduction of Poisson maps and geometric construction of inte-
\linebreak \phantom{fak}\,\,grable systems  via successive reductions
\endsubhead
\bigskip

The goal of this section is to develop a general scheme of constructing
integrable systems based on realization in symmetric coboudary dynamical
Lie algebroids and symmetric coboundary dynamical Poisson groupoids.
In order to do this, we have to consider the method of Dirac reduction.
We begin by recalling the notion of a Dirac submanifold as recently 
introduced in \c{X}. It is a generalization of the notion of cosymplectic
submanifolds of Weinstein \c{W1}. For convenience, we shall formulate
our results in this section in the differentiable category, but it
will be clear that the results are also valid for the holomorphic
category.  

\definition
{Definition 3.1} Let $(P,\pi)$ be a Poisson manifold.  A submanifold $Q$
of $P$ is a Dirac submanifold iff there exists a Whitney sum decomposition
$$T_{Q}P = TQ\oplus V_{Q}\eqno(3.1)$$
where $V_{Q}^{\perp}$ is a Lie subalgebroid of the cotangent Lie
algebroid $T^{*}P$.
\enddefinition

If $Q$ is a Dirac submanifold of $(P,\pi)$, then necessarily $Q$ carries
a natural Poisson structure $\pi_{Q}$ whose symplectic leaves are given
by the intersection of $Q$ with the symplectic leaves of $P$.  Indeed,
$\pi^{\sharp}_{Q}:T^{*}Q\longrightarrow TQ$ is just the anchor map of
the Lie subalgebroid $T^{*}Q\simeq V_{Q}^{\perp}$ of $T^{*}P$. Moreover, 
from the knowledge of the injective Lie algebroid morphism
$T^{*}Q\longrightarrow T^{*}P$, it is easy to show that
$$\pi^{\sharp}_{Q} = pr\circ \pi^{\sharp}\mid_{Q}\circ pr^* \eqno(3.2a)$$
where $pr:T_{Q}P\longrightarrow TQ$ is the projection map induced
by the decomposition in (3.1) and $pr^*$ is its dual.
Alternatively, we have 
$$\pi\mid_{Q} = \pi_{Q} + \pi^{\prime}\eqno(3.2b)$$
where $\pi^{\prime} \in Sect(\wedge^{2}V_{Q}).$
 
We shall call $Q$ equipped with the induced Poisson structure a
{\it Dirac reduction\/} of $P.$   
\smallskip
\noindent{\bf Remark 3.2.} An important class of Dirac submanifolds is 
given by the cosymplectic submanifolds
of Weinstein \c{W1}, in which case $V_{Q} = \pi^{\sharp}\mid_{Q}((TQ)^{\perp}).$
Note that when $P$ is symplectic, the cosymplectic submanifolds
of $P$ are precisely its symplectic submanifolds.  We shall
give another important class of examples in Proposition 3.5 below.
\medskip
Since we will be dealing with realization maps into the dual bundles
of symmetric coboundary dynamical Lie algebroids (resp. symmetric
coboundary dynamical Poisson groupoids), the following result is
fundamental in reducing such maps.

\proclaim
{Theorem 3.3} Let $\phi:P_{1}\longrightarrow P_{2}$ be a Poisson
map and let $Q_1\subset P_1$, $Q_2\subset P_2$ be Dirac submanifolds
with respective Whitney sum decompositions
$$T_{Q_{1}}P_{1} = TQ_{1}\oplus V_{Q_{1}},\quad
  T_{Q_{2}}P_{2} = TQ_{2}\oplus V_{Q_{2}}.$$
Then under the assumptions that\newline
\noindent (i) $\phi(Q_1) \subset Q_2$,
\smallskip
\noindent (ii) $T_{x}\phi\, (V_{Q_1})_{x} \subset
 (V_{Q_2})_{\phi(x)}, \,\,\forall x\in Q_{1}$,
\smallskip
\noindent the map $\phi\mid Q_{1}: Q_{1}\longrightarrow Q_{2}$ is a
Poisson map, when $Q_1$ and $Q_2$ are equipped with the
induced Poisson structures.
\endproclaim

\demo
{Proof} Let $\pi^{\sharp}_{P_{i}}$, $\pi^{\sharp}_{Q_{i}}$ be
the bundle maps associated with the Poisson structures on
$P_{i}$, $Q_{i}$, $i=1,2$.  Since $Q_{1}$ is a Dirac
submanifold of $P_{1}$, we have
$$\pi^{\sharp}_{P_{1}}(x)(\alpha) =\pi^{\sharp}_{Q_{1}}(x)(\alpha\mid T_{x}
   Q_{1}) + \widetilde \pi^{\sharp}_{Q_{1}}(x)(\alpha\mid (V_{Q_{1}})_{x})
   \quad\quad\quad\quad (*)$$
for all $x\in Q_{1}$, $\alpha\in T^{*}_{x}P_{1}$, where 
$\widetilde \pi^{\sharp}_{Q_{1}}:V^{*}_{Q_{1}}\longrightarrow V_{Q_{1}}.$
Similarly,
$$\pi^{\sharp}_{P_{2}}(y)(\beta) =\pi^{\sharp}_{Q_{2}}(y)(\beta\mid T_{y}
   Q_{2}) + \widetilde \pi^{\sharp}_{Q_{2}}(y)(\beta\mid (V_{Q_{2}})_{y})
   \quad\quad\quad\quad (**)$$  
for all $y\in Q_{2}$, $\beta\in T^{*}_{y}P_{2}$, where 
$\widetilde \pi^{\sharp}_{Q_{2}}:V^{*}_{Q_{2}}\longrightarrow V_{Q_{2}}.$
Now, it follows from (*) that
$$
\aligned
       &T_{x}\phi\cdot\pi^{\sharp}_{P_{1}}(x)\cdot T^{*}_{x}\phi(\beta)\\
  =\,\,&T_{x}\phi\cdot\pi^{\sharp}_{Q_{1}}(x)(T^{*}_{x}\phi(\beta)\mid T_{x}
        Q_{1}) + T_{x}\phi\cdot\widetilde \pi^{\sharp}_{Q_{1}}(x)
       (T^{*}_{x}\phi(\beta)\mid (V_{Q_{1}})_{x})
\endaligned
$$
for all $x\in Q_{1}$, $\beta\in T^{*}_{\phi(x)}P_{2}$.  From
assumption (i), we have 
$T_{x}\phi\cdot\pi^{\sharp}_{Q_{1}}(x)(T^{*}_{x}\phi(\beta)\mid T_{x}
  Q_{1})\in T_{\phi(x)}Q_{2}.$  On the other hand, assumption (ii)
implies that
$T_{x}\phi\cdot\widetilde \pi^{\sharp}_{Q_{1}}(x)
 (T^{*}_{x}\phi(\beta)\mid (V_{Q_{1}})_{x})\in (V_{Q_{2}})_{\phi(x)}.$
Since $\phi$ is Poisson, it follows from (**) above that we also have
$$
\aligned
       &T_{x}\phi\cdot\pi^{\sharp}_{P_{1}}(x)\cdot T^{*}_{x}\phi(\beta)\\
 =\,\, &\pi^{\sharp}_{Q_{2}}(\phi(x))(\beta\mid T_{\phi(x)}
   Q_{2}) + \widetilde \pi^{\sharp}_{Q_{2}}(\phi(x))(\beta\mid 
   (V_{Q_{2}})_{\phi(x)})
\endaligned
$$
for all $x\in Q_{1}$, $\beta\in T^{*}_{\phi(x)}P_{2}.$  Therefore,
upon equating the two expressions for
$T_{x}\phi\cdot\pi^{\sharp}_{P_{1}}(x)\cdot T^{*}_{x}\phi(\beta)$,
we obtain 
$$
\aligned
       &T_{x}\phi\cdot\pi^{\sharp}_{Q_{1}}(x)(T^{*}_{x}\phi(\beta)\mid T_{x}
         Q_{1})\\
  =\,\,&\pi^{\sharp}_{Q_{2}}(\phi(x))(\beta\mid T_{\phi(x)}Q_{2})
\endaligned
$$
which shows that $\phi\mid Q_{1}:Q_{1}\longrightarrow Q_{2}$ is
Poisson, as desired.
\pf
\enddemo

\definition
{Definition 3.4} The map $\phi\mid Q_{1}: Q_{1}\longrightarrow Q_{2}$
in the theorem above will be called a {\it Dirac reduction\/} of the
Poisson map $\phi:P_{1}\longrightarrow P_{2}$.
\enddefinition

The following result gives an important class of Dirac submanifolds
which plays a key role in this work.

\proclaim
{Proposition 3.5 \c{X}} Let $\sigma:P\longrightarrow P$ be a Poisson
involution, i.e., an involution which is also a Poisson map.  Then 
its stable locus $Q$ is a Dirac submanifold 
of $P$ with $V_{Q} = \bigcup_{x\in Q}\, ker(T_{x}\sigma + 1).$
\endproclaim

As a consequence of this result, the stable locus of a Poisson
involution carries a natural Poisson structure. This fact was
also noted in \c{FP} and was implicit in the earlier work of several 
authors \c{Bon}, \c{Boal}. (See also p.194 of \c{RSTS}.)

In the special case when the Dirac submanifolds in Theorem 3.3 are the stable
loci of Poisson involutions, we have the following result.

\proclaim
{Corollary 3.6} Let $\sigma_{1}:P_{1}\longrightarrow P_{1}$,
$\sigma_{2}:P_{2}\longrightarrow P_{2}$ be Poisson involutions
with stable loci given by $Q_1$ and $Q_2$ respectively.  If
$\phi:P_{1}\longrightarrow P_{2}$ is a Poisson map which commutes
with $\sigma_1$, $\sigma_2$,i.e. $\sigma_{2}\circ\phi = \phi\circ \sigma_{1}$,
then
$\phi\mid Q_{1}:Q_{1}\longrightarrow Q_{2}$ 
is a
Poisson map, when $Q_{1}$ and $Q_{2}$ are equipped with
the induced structures.
\endproclaim

\demo
{Proof} Under the assumption that $\phi$ commutes with the Poisson
involutions, it is easy the check that the conditions in Theorem 3.3
are satisfied with 
$$V_{Q_{1}} = \bigcup_{x\in Q_{1}}\, ker(T_{x}\sigma_{1} + 1),\quad
  V_{Q_{2}} = \bigcup_{y\in Q_{2}}\, ker(T_{y}\sigma_{2} + 1).$$
Hence the assertion follows. \pf
\enddemo

We next consider the problem of reducing a Hamiltonian $G$-space
$P$ to a Dirac submanifold $Q$. Note that $Q$ is in general not a
$G$-space.  So a  natural question is: under
what condition is $Q$ a Hamiltonian $H$-space for some Lie
subgroup $H\subset G$ ? 

\proclaim
{Proposition 3.7} Let $\Phi:G\times P\longrightarrow P$
be a Hamiltonian group action of $G$ on the Poisson manifold
$(P, \pi)$, and let $Q$ be a Dirac submanifold of
$P$ with Whitney sum decomposition $T_{Q}P = TQ\oplus V_{Q}$.  
If $H$ is a Lie subgroup of 
$G$ with $Lie(H) = \fh$ and if the action $\Phi$ induces an action of $H$ on 
$Q$ satisfying
$$T_{x}\Phi_{h}(V_x) \subset V_{\Phi_{h}(x)}, \quad \forall h\in H, x\in Q,$$
where 
$V_{Q} = \bigcup_{q\in Q} V_{q},$ then the $H$-action 
on $Q$ is also a Hamiltonian group action.
Moreover, if $J:P\longrightarrow \fg^{*}$ is a $G$-equivariant
momentum map for $\Phi$, then the map
$J_{Q}= i^{*}\circ (J\mid Q):Q\longrightarrow \fh^{*}$ is
a $H$-equivariant momentum map for the $H$-action on $Q$.  Here, 
$i^*$ is the dual map of the Lie inclusion $i:\fh\longrightarrow \fg.$ 
\endproclaim

\demo
{Proof} For $h\in H$, the assertion that 
$\Phi_{h}\mid Q:Q\longrightarrow Q$ is Poisson is a consequence of
Theorem 3.3.  To show that $J_Q$ is an equivariant
momentum map for the $H$ action on $Q$, note that
$${d\over dt}\Big|_{t=0} \Phi_{e^{tZ}}(x) = \pi^{\sharp}_{Q}(x)
  (d\widehat{J}(Z)(x)\mid T_{x}Q)$$
for all $Z\in \fh$ and $x\in Q$. But it is easy to check that
the map $\widehat{J}_{Q}:\fh\longrightarrow C^{\infty}(Q)$ defined by
$\widehat{J}_{Q}(Z)(x) = <J_{Q}(x),Z>$, $Z\in \fh$, 
$x\in Q$ satisfies 
$d\widehat{J}_{Q}(x) = d\widehat{J}(Z)(x)\mid T_{x}Q$.  This completes
the proof.
\pf
\enddemo

\proclaim
{Corollary 3.8} Let $\Phi:G\times P\longrightarrow P$
be a Hamiltonian group action of $G$ on the Poisson manifold
$(P, \pi)$ and suppose $\sigma:P\longrightarrow P$ is a
Poisson involution.  If $H$ is a Lie subgroup of $G$
such that 
$$\Phi_{h}\circ \sigma = \sigma\circ \Phi_{h},\quad\forall h\in H,$$
then $\Phi$ induces a Hamiltonian group action of $H$ on the
stable locus $P^{\sigma}$.  Moreover, if 
$J:P\longrightarrow \fg^{*}$ is a $G$-equivariant
momentum map for $\Phi$, then the map
${\widetilde{J}}=i^{*}\circ (J\mid P^{\sigma}):P^{\sigma}\longrightarrow 
\fh^{*}$ is
a $H$-equivariant momentum map for the $H$-action on $P^{\sigma}.$
\endproclaim

\demo
{Proof} It follows from the assumption 
$\Phi_{h}\circ \sigma = \sigma\circ \Phi_{h}, \, h\in H$
that $\Phi$ induces an action of $H$ on $Q =P^{\sigma}$.  Let
$V_{Q} = \bigcup_{x\in Q}\, ker(T_{x}\sigma + 1)$,
we want to show that 
$T_{x}\Phi_{h}(V_x) \subset V_{\Phi_{h}(x)}, \,  h\in H, x\in Q.$
For this purpose, take any $v\in V_{x}.$  Then we have
$$\aligned
         &(T_{\Phi_{h}(x)}\sigma + 1) T_{x}\Phi_{h}(v)\\
      = \, &T_{x}(\Phi_{h}\circ\sigma)(v) + T_{x}\Phi_{h}(v)\quad
          ({\hbox{since}}\,\, \Phi_{h}\circ \sigma = \sigma\circ \Phi_{h})\\
      = \, &0
\endaligned
$$
where we have used the property $T_{x}\sigma (v) = -v$ in the last step.
Hence the assertion follows from Proposition 3.7.
\pf
\enddemo

We next discuss Poisson involutions on $(A\Gamma,\{\cdot,\cdot\}_{A\Gamma})$,
where $\{\cdot,\cdot\}_{A\Gamma}$ is the Lie-Poisson structure in
(2.11). The following result was obtained in \c{X} using  Lie bialgebroid
theory \c{MX}.   We will give an elementary proof based on the explicit
formula in (2.11). 

\proclaim
{Proposition 3.9} Let $s:\fg \longrightarrow \fg$ be an involutive
Lie algebra anti-morphism which preserves $\fh$ and assume that
$s\circ R(q)\circ s^{*} = - R(s^{*}_{\fh} q)$ for all $q\in U$
where $s_{\fh} = s\mid_{\fh}.$  Then the map
$$\sigma:(A\Gamma,\{\cdot,\cdot\}_{A\Gamma})\longrightarrow
  (A\Gamma,\{\cdot,\cdot\}_{A\Gamma}), (q,\lambda,X)\mapsto
  (s^{*}_{\fh}(q), -s^{*}_{\fh}(\lambda), s(X))\eqno(3.3)$$
is a Poisson involution.
\endproclaim

\demo
{Proof} From the property that $s\circ R(q)\circ s^{*} = - R(s^{*}_{\fh} q)$,
it follows that
$s(dR(q)(\lambda)s^{*}\xi) = -dR(s^{*}_{\fh} q)(s^{*}_{\fh}\lambda)(\xi).$
The rest of the proof is plain.
\pf
\enddemo

\noindent{\bf Remark 3.10.} The virtue of our direct verification in
the above proof lies in the fact that it extends to more general
constructions in which the Lie bialgebroid structure is lost.
\medskip
We shall call $(A^{*}\Gamma, [\cdot,\cdot]_{A^{*}\Gamma},a_{*},\sigma^{*})$
a {\it symmetric coboundary dynamical Lie algebroid\/}.
In order to compute the Poisson structure on the stable locus,
we shall introduce some notation which we shall use
in the rest of the section.  Let $(P, \{\cdot,\cdot\}_P)$ be a
Poisson manifold and suppose $\tau:P\longrightarrow P$
is a Poisson involution with stable locus $P^{\tau}$.  Then
for $\varphi \in C^{\infty}(P)$, we put 
$\widetilde \varphi = \varphi\mid P^{\tau}$ and
$\varphi^{\tau} = {1\over 2} (\varphi + \tau^{*}\varphi).$
Since for $Q = P^{\tau}$, we have 
$V_{Q} = \bigcup_{x\in Q}\, ker(T_{x}\tau + 1)$ in the
Whitney sum decomposition for $T_{Q}P$. Hence it follows from (3.2) that
the induced Poisson structure on $P^{\tau}$ is given by the
formula
$$\{\varphit,\psit\}_{P^{\tau}}(x) = 
   \{\varphi^{\tau},\psi^{\tau}\}_{P}(x)\eqno(3.4)$$
for $x\in P^{\tau}$ and  $\varphi$, $\psi\in C^{\infty}(P).$

\proclaim
{Proposition 3.11}  The Poisson structure on the stable locus
$A\Gamma^{\sigma}$ of the Poisson involution in (3.3) is
given by
$$\eqalign{& \{\TF_1, \TF_2\}_{{A\Gamma}^{\sigma}} (q,\lambda,X)\cr
  =\,&-<\lambda, [\delta_{2}\TF_1,\delta_{2}\TF_2]>+
     <dR(q)(\lambda)\delta \TF_1,\delta\TF_2>\cr
     & +<X, -ad^{*}_{R(q)\delta \TF_1-\delta_{2}\TF_1}
          \delta \TF_2 + ad^{*}_{R(q)\delta \TF_2-\delta_{2}\TF_2}
          \delta \TF_1 >\cr
     &-<q,[\delta_{2}\TF_1,\delta_{1}\TF_2]+[\delta_{1}\TF_1,
       \delta_{2}\TF_2]>+<\delta_{1}\TF_2, \iota^{*}\delta \TF_1>
       -<\delta_{1}\TF_1,\iota^{*}\delta\TF_2>\cr}\eqno(3.5)$$
for $F_1, F_2 \in C^{\infty}(A\Gamma)$, 
$(q,\lambda, X)\in {A\Gamma}^{\sigma}$ where
$$\eqalign {& \delta_{1} \TF_i := {1\over 2}(\delta_{1} F_i +
   s_{\fh}(\delta_{1} F_i)), \quad \delta_{2}\TF_i :=
   {1\over 2}(\delta_{2}\TF_i - s_{\fh}(\delta_{2} \TF_i)),\cr
   & \delta \TF_i := {1\over 2} (\delta \TF_i +
     s^{*}(\delta \TF_i)),\,\,\,\,i=1,2.\cr}\eqno(3.6)$$
Moreover, the Hamiltonian vector
field on $A\Gamma^{\sigma}$ generated by $\TF$ is of the form
$$\eqalign{
          & X_{\TF}(q,\lambda,X)\cr
       =\,  & (\iota^{*}\delta\TF - ad^{*}_{\delta_{2}\TF} q,
             -ad^{*}_{\delta_{2}\TF}\lambda + \iota^{*}ad^{*}_{X}\delta\TF
           -ad^{*}_{\delta_{1}\TF}q,\cr
          & [X, R(q)\delta\TF-\delta_{2}\TF] + dR(q)(\lambda)\delta\TF
            -\delta_{1}\TF + R(q)ad^{*}_{X}\delta\TF).\cr}\eqno(3.7)$$
\endproclaim

\demo
{Proof} The bundle map $\pi^{\sharp}$ corresponding to the
Poisson bracket $\{\cdot,\cdot\}_{A\Gamma}$ in (2.11) is given
by
$$
\aligned
       & \pi^{\sharp}(q,\lambda, X)(Z_{1},Z_{2},\xi)\\
     = &(\iota^{*}\xi - ad^{*}_{Z_{2}} q,
         -ad^{*}_{Z_{2}} \lambda
         +\iota^{*}ad^{*}_{X} \xi
         -ad^{*}_{Z_{1}} q,\\
       & \quad
         [X, R(q)\xi - Z_{2}]
         + dR(q)(\lambda) \xi - Z_{1}
         +R(q)(ad^{*}_{X} \xi))
\endaligned
$$
where $(q,\lambda, X)\in A\Gamma$ and 
$(Z_{1}, Z_{2}, \xi) \in \fh\times \fh\times \fg^{*}\simeq
 T^{*}_{(q,\lambda, X)} (A\Gamma).$
Let $Q = A\Gamma^{\sigma}$ and let 
$\iota_{Q}:Q\longrightarrow A\Gamma$ be the canonical inclusion.
In this case, the bundle $V_{Q}$ in the vector bundle decomposition
$T(A\Gamma) = TQ \oplus V_{Q}$ is just the bundle of $-1$ eigenspaces
of $T\sigma$.  Therefore, if $(q,\lambda, X)\in Q$, and
$(Z_{1}, Z_{2}, \xi) \in \fh\times \fh\times \fg^{*}\simeq
 T^{*}_{(q,\lambda, X)} (A\Gamma),$ it follows that the
bundle map of the induced Poisson structure on $Q$ is given by
$$
\aligned
       & \pi^{\sharp}_{Q}(q,\lambda, X)(T^{*}_{(q,\lambda,X)} \iota_{Q}
         (Z_{1},Z_{2},\xi))\\
     = &(\iota^{*}\widetilde \xi - ad^{*}_{\widetilde Z_{2}} q,
         -ad^{*}_{\widetilde Z_{2}} \lambda
         +\iota^{*}ad^{*}_{X} \widetilde \xi
         -ad^{*}_{\widetilde Z_{1}} q,\\
       & \quad
         [X, R(q)\widetilde \xi - \widetilde Z_{2}]
         + dR(q)(\lambda)\widetilde \xi - \widetilde Z_{1}
         +R(q)(ad^{*}_{X}\widetilde \xi))
\endaligned
$$
where 
$$
\aligned
       & \widetilde Z_{1} = {1\over 2} (Z_{1} + s_{\fh}(Z_{1})), \quad
         \widetilde Z_{2} = {1\over 2} (Z_{2} - s_{\fh}(Z_{2})), \\
       & \widetilde \xi = {1\over 2} (\xi + s^{*}(\xi)).
\endaligned
$$
Since 
$d \TF (q,\lambda,X) = T^{*}_{(q,\lambda,X)} \iota_{Q}
 d F (q,\lambda, X)$, the formula for the vector field
$X_{\TF}$ is immediate  from the above expression.
On the other hand, the formula for the Poisson bracket is a
consequence of (2.3) and (3.4) as we have 
$d F^{\sigma}_{i} (q,\lambda, X) = (\delta_{1}\widetilde F_{i},
\delta_{2} F_{i}, \delta \widetilde F_{i})\,\,, i = 1,2.$
\pf
\enddemo

We now turn to corresponding results for the coboundary dynamical
Poisson groupoids.  The following result was also obtained in
\c{X} by invoking Lie bialgebroid theory.  We can, of course,
verify the assertion in a direct way by using the formula
in (2.7).

\proclaim
{Proposition 3.12} Let $R$ be an $H$-equivariant
classical dynamical r-matrix such that
$s\circ R(q)\circ s^{*} = - R(s^{*}_{\fh} q)$
for all $q\in U$ where $s$ is as in Proposition 3.9.
If $(\Gamma, \{\, \cdot, \cdot \,\}_{R})$ is the
coboundary dynamical Poisson groupoid associated to $R$,
then the map
$$\Sigma:(\Gamma, \{\, \cdot, \cdot \,\}_{R})\longrightarrow
  (\Gamma, \{\, \cdot, \cdot \,\}_{R}), (u,g,v)\mapsto
  (s^{*}_{\fh}(v),S(g),s^{*}_{\fh}(u))\eqno(3.8)$$
is a Poisson involution, where $S:G\longrightarrow G$ is
the group anti-morphism which integrates $s$.
\endproclaim
We shall call $(\Gamma, \{\cdot,\cdot\}_{R},\Sigma)$ a
{\it symmetric coboundary dynamical Poisson groupoid\/}.

\proclaim
{Proposition 3.13} With the involution $\Sigma$ in (3.8), the induced
Poisson structure on its stable locus $\Gamma^{\Sigma}$ is given
by 
$$\eqalign {&\{\varphit,\psit\}_{\Gamma^{\Sigma}} (u,g,s^{*}_{\fh}(u))\cr
  = & \,\, 2<u, [\delta_1\varphit,\,\delta_1\psit]> - 
   2<\iota\delta_1 \varphit,\, D\psit> +
    2 <\iota\delta_1 \psit,\, D\varphit>\cr 
       & +<R(s^{*}_{\fh}(u)) D'\varphit, D'\psit>- <R(u) D\varphit, D\psit>\cr
=& \,\,2<u, [\delta_1\varphit,\,\delta_1\psit]> - 2<\iota\delta_1 \varphit,\, 
   D\psit>
 +2 <\iota\delta_1 \psit,\, D\varphit>\cr
   &  -2<R(u) D\varphit, D\psit>\cr}\eqno(3.9)$$
for $\varphi, \psi \in C^{\infty}(\Gamma)$, $(u,g,s^{*}_{\fh}(u))\in 
\Gamma^{\Sigma}$,
where 
$$\eqalign{
   &\delta_{1}\varphit:= {1\over 2}(\delta_{1}\varphi + 
     s_{\fh}(\delta_{2}\varphi)),
        \quad D\varphit:={1\over 2}(D\varphi + s^*(D'\varphi)),\cr
   & D'\varphit:={1\over 2}(D'\varphi + s^*(D\varphi)),\cr}\eqno(3.10)$$
and similarly for $\psit.$  Here, $\delta_{1}\varphi$,$\delta_{2}\varphi$
are the partial derivatives of $\varphi$ with respect to the variables in
$U$ and $D'\varphi$, $D\varphi$ are the
left and right gradients of $\varphi$ with respect to the variable
in $G$.  Hence the Hamiltonian vector field on $\Gamma^{\Sigma}$ 
generated by $\varphit$ is of the form
$$\eqalign{
         & X_{\varphit} (u,g,s^{*}_{\fh}(u))\cr
      =  &(ad^{*}_{\delta_{1}\varphit} u + \iota^{*}D\varphit, 
          -T_{e}r_{g}\delta_{1}\varphit -T_{e}l_{g}s^{*}_{\fh}
          (\delta_{1}\varphit)
           +T_{e}l_{g}R(s^{*}_{\fh}(u))s^{*}(D\varphit)\cr
         &-T_{e}r_{g}R(u)D\varphit,
         s^{*}_{\fh}(ad^{*}_{\delta_{1}\varphit} u + \iota^{*}D\varphit)).\cr}
         \eqno(3.11)$$
\endproclaim

\demo
{Proof} According to (2.7) and (3.4), we can express 
the bracket $\{\varphit,\psit\}_{\Gamma^{\Sigma}}$ in the following
form:
$$\eqalign {\{\varphit,\psit\}_{\Gamma^{\Sigma}} (u,g,s^{*}_{\fh}(u))=
  & <u, [\delta_1\varphit,\,
  \delta_1\psit]> -<s^{*}_{\fh}(u), [\delta_2\varphit,\, \delta_2\psit]>\cr
& - <\iota\delta_1 \varphit,\, D\psit> - <\iota\delta_2 \varphit, D'\psit>\cr
& + <\iota\delta_1 \psit,\, D\varphit> + <\iota\delta_2 \psit , D'\varphit>\cr
& +<R(s^{*}_{\fh}(u)) D'\varphit, D'\psit> - <R(u) D\varphit, D\psit>\cr}
$$
where $\delta_{2}\varphit:= {1\over 2}(\delta_{2}\varphi + 
 s_{\fh}(\delta_{1}\varphi))$
and the other derivatives are defined in (3.10).        
However, it is easy to show that
$$<s^{*}_{\fh}(u), [\delta_2\varphit,\, \delta_2\psit]>
= -<u, [\delta_1\varphit,\,\delta_1\psit]>,\,\,
<\iota\delta_2 \varphit, D'\psit> =<\iota\delta_1 \varphit,\, D\psit>$$
and \,\,$<R(s^{*}_{\fh}(u)) D'\varphit, D'\psit>= - <R(u) D\varphit, D\psit>.$
Therefore, the above expression for the bracket simplifies to the ones
in the statement of the theorem. The computation of the vector field
proceeds as in the proof of Proposition 3.11 and so we skip the details.
\pf 
\enddemo

Now let us recall from \c{L2} that the Lie-Poisson structure on
the dual bundle $A^{*}\Gamma$ of the trivial Lie algebroid
$A\Gamma$ is given by
$\{{\Cal F}, {\Cal G}\}_{A^{*}\Gamma}(q,p,\xi) =\,
  <\delta_{2}{\Cal F},\delta_{1}{\Cal G}>
-<\delta_{1}{\Cal F}, \delta_{2}{\Cal G}>+ 
  <\xi,[\delta{\Cal F},\delta{\Cal G}]>.$

We shall leave the proof of the next result to the reader.

\proclaim
{Proposition 3.14} Let $b:\fh\longrightarrow \fh$ be an involutive
linear map and suppose $c:\fg\longrightarrow \fg$ is an involutive
Lie algebra morphism.  Then the map
$$\theta:(A^*\Gamma, \group_{A^*\Gamma})\longrightarrow
  (A^*\Gamma, \group_{A^*\Gamma}), (q,p, \xi)\mapsto 
  (b^*(q), b(p), c^*(\xi))\eqno(3.12)$$
is a Poisson involution.  Moreover, the induced Poisson
structure on the stable locus $A^{*}\Gamma^{\theta}$ is 
given by
$$\eqalign{
       &\{\widetilde{\Cal F}, \widetilde{\Cal G}\}_{A^{*}\Gamma^{\theta}}
         (q,p,\xi)\cr
   =\,\,&<\delta_{2}\widetilde{\Cal F},\delta_{1}\widetilde{\Cal G}>
     -<\delta_{1}\widetilde{\Cal F}, \delta_{2}\widetilde{\Cal G}>+ 
      <\xi,[\delta\widetilde{\Cal F},\delta\widetilde{\Cal G}]>\cr}\eqno(3.13)
$$
for ${\Cal F}, {\Cal G} \in C^{\infty}(A^*\Gamma)$ and
$(q,p,\xi)\in A^*\Gamma^{\theta}$ where
$$\eqalign{
       &\delta_{1}\widetilde{\Cal F}:= {1\over 2}(\delta_{1}{\Cal F}
         +b(\delta_{1}{\Cal F})), \quad
        \delta_{2}\widetilde{\Cal F}:= {1\over 2}(\delta_{2}{\Cal F}
         +b^*(\delta_{2}{\Cal F})),\cr
       & \delta\widetilde{\Cal F}:= {1\over 2}(\delta{\Cal F}
         +c(\delta{\Cal F})),\cr}\eqno(3.14)$$
and similarly for $\widetilde{\Cal G}$.
\endproclaim

Thus the induced structure
$\{\cdot,\cdot\}_{A^{*}\Gamma^{\theta}}$ is still a product structure.
Indeed, under the natural isomorphism between $(\fg^{*})^{c^*}$ and
$(\fg^{c})^*$, we can identify the bracket on the stable locus 
$(\fg^{*})^{c^*}$ with the Lie-Poisson structure on $(\fg^{c})^*$.

We are now ready to formulate the main results of this section.
In what follows, let $X$ be a Hamiltonian $H$-space (the $H$-action
will be denoted by ${\Cal C}$) with equivariant
momentum map $J:X\longrightarrow \fg^{*}$, and let 
$\kappa :X\longrightarrow X$  
be a Poisson involution on $X$.
Beginning with $H$-invariant Hamiltonian systems on $X$ which admit
either a realization in 
$(A\Gamma,\{\cdot,\cdot\}_{A\Gamma})$ or
$(\Gamma, \{\, \cdot, \cdot \,\}_{R})$,
we shall show how reduction to Dirac submanifolds followed by Poisson
reduction can lead us to integrable systems.
\medskip
{\bf Case 1. The case of realization in $(A\Gamma,\{\cdot,\cdot\}_{A\Gamma})$}
\medskip
Let $\rho:X\longrightarrow A\Gamma$ be a realization of the Poisson
manifold $X$ in the dual bundle $A\Gamma$ of the Lie algebroid
$A^{*}\Gamma$; i.e., $\rho$ is a Poisson map.  Let us recall 
from Theorem 2.2 that
$A\Gamma$ with the action
$${\Cal A}:H \times A\Gamma \longrightarrow A\Gamma,\,\,
  {\Cal A}_{h} (q,\lambda,\xi) = 
  (Ad^{*}_{h^{-1}}q, Ad^{*}_{h^{-1}}\lambda,Ad_{h}\xi)\eqno(3.15)$$
is a Hamiltonian $H$-space with equivariant momentum map
$$\gamma:A\Gamma\longrightarrow \fh^*,\,\,(q, \lambda, \xi)\mapsto
  \lambda.\eqno(3.16)$$

We begin by making the following assumption:
\medskip
 
\noindent A1. there exists a Poisson involution 
$$\sigma:(A\Gamma,\{\cdot,\cdot\}_{A\Gamma})\longrightarrow
  (A\Gamma,\{\cdot,\cdot\}_{A\Gamma}), (q,\lambda,\xi)\mapsto
  (s^{*}_{\fh}(q), -s^{*}_{\fh}(\lambda), s(\xi))\eqno(3.17)$$
on $A\Gamma$ (where $s$ satisfies the assumptions in Proposition
3.9) such that
$$\sigma\circ \rho = \rho\circ \kappa.\eqno(3.18)$$\newline

Then according to Corollary 3.6, the map $\rho$ restricts
to a Poisson map $\widetilde{\rho}:X^{\kappa}\longrightarrow A\Gamma^{\sigma}$,
when $X^{\kappa}$ and $A\Gamma^{\sigma}$ are equipped with the induced
structures.  Thus the stable locus $X^{\kappa}$ admits a realization
in $A\Gamma^{\sigma}\simeq U_{s}\times \fh^{*}_{s}\times \fg^{s}$, where
$$U_{s} = \{q\in U\mid s^{*}_{\fh}(q) = q\}, \eqno(3.19)$$
$$\fh^{*}_{s} = \{\lambda \in \fh^{*}\mid s^{*}_{\fh}(\lambda) = 
  -\lambda\},\eqno(3.20)$$
and $\fg^{s}$ is the fixed point set of $s.$  Let $I(\fg)$ be
the ring of ad-invariant functions on $\fg$, and let
$I(\fg^{s})$ consists of the restrictions of functions
in $I(\fg)$ to $\fg^{s}.$  If $Pr_{3}$ denote the projection
map from $A\Gamma^{\sigma}\simeq U_{s}\times \fh^{*}_{s} \times \fg^{s}$
to the factor $\fg^{s}$, then a natural family of invariant
functions on $A\Gamma^{\sigma}$ is
$Pr^{*}_{3} I(\fg^{s})$.  Our first result on Poisson commuting functions
will have application in Section 4.2 below.

\proclaim
{Theorem 3.15} Let $\sigma$ be a Poisson involution of the form in (3.17)
on $A\Gamma$ (where $s$ satifies the assumptions in Proposition 3.9) and
suppose  $\fh^{*}_{s}= \{0\}$, then the functions
in $Pr^{*}_{3} I(\fg^{s})$ Poisson commute in $A\Gamma^{\sigma}$.
Consequently, if we assume in addition that A1 is valid, then 
${\widetilde{\rho}}^{*}Pr^{*}_{3} I(\fg^{s})$ is
a Poisson commuting family of functions on $X^{\kappa}$, where
$\widetilde{\rho}=\rho\mid X^{\kappa}$.
\endproclaim

\demo
{Proof} Let $f_1$, $f_2\in I(\fg)$, and let  $\widetilde{f}_1$,
$\widetilde{f}_2$  be their restrictions to $I(\fg^{s})$. Then
from (3.5), we have 
$$\aligned
   &\{Pr^{*}_{3}\widetilde{f}_1,Pr^{*}_{3}\widetilde{f}_2\}_{{A\Gamma}^{
     \sigma}}(q,0, \xi)\\
 =\,& \{\widetilde{Pr^{*}_{3}f_1}, \widetilde{Pr^{*}_{3}f_2}\}_{{A\Gamma}^{
     \sigma}}(q,0,\xi)\\
=\, &<\xi, -ad^{*}_{R(q)\delta {\widetilde{Pr^{*}_{3}f_1}}}
          \delta \widetilde{Pr^{*}_{3}f_2} + ad^{*}_{R(q)\delta
          \widetilde{Pr^{*}_{3}f_2}}
          \delta {\widetilde{Pr^{*}_{3}f_1}}>
\endaligned
$$
where in the last two lines, we have used the same symbol
$Pr_{3}$ to denote the projection map from $A\Gamma$ to $\fg$
and $\widetilde{Pr^{*}_{3}f_i}$ denote the restriction of
$Pr^{*}_{3}f_{i}$ to $A\Gamma^{\sigma}, \,\,i=1,2.$
Now, by direct calculation, we find
$$ \delta {\widetilde{Pr^{*}_{3}f_i}}={1\over 2}(df_i+s^*(df_i)),\,\,i=1,2.$$
Therefore, upon substituting into the above expression, we obtain
$$\aligned
   &\{Pr^{*}_{3}\widetilde{f}_1,Pr^{*}_{3}\widetilde{f}_2\}_{{A\Gamma}^{
     \sigma}}(q,0, \xi)\\
=\,&{1\over 4}<[\xi,R(q)(df_1+s^{*}(df_1))], df_2+s^{*}(df_2)> -
(1\leftrightarrow 2)\\
=\,&{1\over 4}<R(q)(df_1+s^{*}(df_1)), ad^{*}_{\xi}df_2 + ad^{*}_{\xi}
    s^{*}(df_2)> - (1\leftrightarrow 2).
\endaligned
$$
But as $\xi\in \fg^{s}$, we have 
$ad^{*}_{\xi}\circ s^{*} = -s^{*}\circ ad^{*}_{\xi}.$
Hence the first assertion follows from the fact that
$ad^{*}_{\xi}df_i = 0$, $i=1,2.$
The second assertion is now clear as assumption A1 implies that
$\widetilde{\rho}$ is Poisson by Corollary 3.6.
\pf
\enddemo

In the general case when $\fh^{*}_{s}\neq \{0\}$, the functions in
$Pr^{*}_{3} I(\fg^{s})$ is no longer a Poisson commuting family
on $A\Gamma^{\sigma}$.  Indeed, by a computation similar to the
one in the proof of the above theorem, we have
$$\eqalign{
  &\{Pr^{*}_{3}\widetilde{f}_1,Pr^{*}_{3}\widetilde{f}_2\}_{{A\Gamma}^{
     \sigma}}(q,\lambda, \xi)\cr
=\,&{1\over 4}<dR(q)(\lambda)(df_1+s^{*}(df_1)),df_2+s^{*}(df_2)>\cr}
\eqno(3.21)$$
for $f_1,f_2\in I(\fg).$  Nevertheless, it is clear from this
expression that when we restrict to the submanifold
$U_{s}\times \{0\}\times \fg^{s}$ of $A\Gamma^{\sigma}$, the bracket
vanishes.  Note that in general neither $X^{\kappa}$ nor $A\Gamma^{\sigma}$
are Hamiltonian $H$-spaces.  We now discuss a situation where
we can obtain Poisson commuting functions on a reduced phase space.
Motivated by our application in Section 4.1 below, we shall make the
following assumptions to prepare the way for Poisson reduction:
\medskip
\noindent A2. the realization map $\rho$ is $H$-equivariant,\newline
\noindent A3. for some Lie subgroup $D$ of $H$,
$${\Cal A}_{d}\circ \sigma = \sigma \circ {\Cal A}_{d},\quad
  {\Cal C}_{d}\circ \kappa = \kappa \circ {\Cal C}_{d},\quad 
  \forall d\in D,\eqno(3.22)$$\newline
\noindent A4. if $\frak d = Lie (D)$ and 
$\fh^{*}_{s}$ is as in (3.20),
we assume 
$$\fh^{*}_{s} \subset \frak d^{*}.\eqno(3.23)$$

\proclaim
{Proposition 3.16} Under assumptions A1-A4, the stable loci
$X^{\kappa}$, $A\Gamma^{\sigma}$ are Hamiltonian $D$-spaces
with equivariant momentum maps
${\widetilde{J}} = i^{*}_{\frak d}\circ (J\mid X^{\kappa})$
and ${\widetilde{\gamma}}= \gamma\mid A\Gamma^{\sigma}$ respectively,
where $i_{\frak d}:\frak d\longrightarrow \fh$ in the natural 
inclusion and
$i^{*}_{\frak d}$ is its dual map.
Moreover, the map
$${\widetilde{\rho}}= \rho\mid X^{\kappa}:X^{\kappa}\longrightarrow
  A\Gamma^{\sigma}\eqno(3.24)$$
is a $D$-equivariant Poisson map.  
\endproclaim

\demo
{Proof} From A3 and Corollary 3.8, it follows that the actions
${\Cal A}$ and ${\Cal C}$ induce Hamiltonian group actions of
$D$ on the stable loci $A\Gamma^{\sigma}$ and $X^{\kappa}$
respectively.  Using the second part of the same corollary and
A4, we can easily obtain the equivariant momentum maps of
these induced actions.  Finally, that the map ${\widetilde{\rho}}$
is well-defined and Poisson is a consequence of A1 and Corollary 3.6,
and its $D$-equivariance is obvious from A2. 
\pf
\enddemo

The above proposition completes the first stage of our reduction
process in the general case and prepares the way for Poisson
reduction.  In order to obtain Poisson commuting functions in
this general case, 
we shall make an additional assumption:
\medskip
\noindent A5. for some regular value $\mu\in \frak d^{*}$ of 
${\widetilde{J}},$
$${\widetilde{\rho}}({\widetilde{J}}^{-1}(\mu))\subset
 {\widetilde{\gamma}}^{-1}(0)= U_{s}\times \{0\}\times \fg^{s}.\eqno(3.25)$$

With this additional assumption, we will construct the integrable
systems and their realizations by Poisson reduction of the map
$\widetilde{\rho}$ in Proposition 3.16.  Let
$D_{\mu}$ be the isotropy subgroup of $D$ for the $D$-action on 
$X^{\kappa}$, then by Poisson reduction \c{MR},\c{OR}, the
variety $X^{\kappa}_{\mu} = {\widetilde{J}}^{-1}(\mu)/D_{\mu}$
inherits a unique Poisson structure $\{\,\cdot,\cdot \,\}_{X^{\kappa}_{\mu}}$
satisfying 
$$\pi_{\mu}^* \{f_1, f_2 \}_{X^{\kappa}_{\mu}} = i_{\mu}^* \{ {f^{\prime}_1},
{f^{\prime}_2} \}_{X^{\kappa}} .\eqno(3.26)$$
Here, $i_{\mu}:{\widetilde{J}}^{-1} (\mu) \longrightarrow X^{\kappa}$ is the 
inclusion map,
$\pi_{\mu}:{\widetilde{J}}^{-1} (\mu)\longrightarrow X^{\kappa}_{\mu}$ is 
the canonical projection,
$f_1$, $f_2 \in C^{\infty} (X^{\kappa}_{\mu})$, and ${f^{\prime}_1}$,
${f^{\prime}_2}$ are (locally defined) smooth extensions of 
$\pi_{\mu}^* f_1$, $\pi_{\mu}^* f_2$ with differentials vanishing
on the tangent spaces of the $D$-orbits. Similarly, we have the Poisson 
variety
$$\left(A\Gamma^{\sigma}_0 = {\widetilde{\gamma}}^{-1}(0)/ D, 
\{\,\cdot,\cdot \,\}_{A\Gamma^{\sigma}_0}\right),
\eqno(3.27)$$
with the inclusion map $i_D  :{\widetilde{\gamma}}^{-1}(0) \longrightarrow 
A\Gamma^{\sigma}$ and
the canonical projection $\pi_D:{\widetilde{\gamma}}^{-1}(0) \longrightarrow 
A\Gamma^{\sigma}_0.$
If $Pr_{i}$ denotes the projection map onto the $i$-th factor of 
$U_{s}\times \fh^{*}_{s}\times \fg^{s} \simeq A\Gamma$,
$i=1,2,3,$ we put
$$m = Pr_{1}\circ {\widetilde{\rho}}: X^{\kappa} \longrightarrow U_{s}, 
\eqno(3.28)$$
$$\tau = Pr_{2}\circ {\widetilde{\rho}}: X^{\kappa} \longrightarrow \fh^{*}_{s},
\eqno(3.29)$$
$$L = Pr_{3}\circ {\widetilde{\rho}}:X^{\kappa}\longrightarrow \fg^{s}.
\eqno(3.30)$$
Clearly, functions in $i_D^* Pr_3^* I(\fg^{s}) \subset 
C^{\infty} ({\widetilde{\gamma}}^{-1}(0))$
are $D$-invariant, hence they descend to functions in 
$C^{\infty} (A\Gamma^{\sigma}_0)$.  On the other hand, it follows from
Proposition 3.16 that the functions in $i_{\mu}^* L^* I(\fg^{s}) 
\subset
C^{\infty} ({\widetilde{J}}^{-1}(\mu))$ drop down to functions in 
$C^{\infty} (X^{\kappa}_{\mu}).$  
Now,by Proposition 3.16 and assumption A5, it follows from 
Theorem 2.14 of \c{OR} 
that ${\widetilde{\rho}}$ induces a unique Poisson map (called the
reduction of $\widetilde{\rho}$)
$$\widehat{\rho}: X^{\kappa}_{\mu} \longrightarrow A\Gamma^{\sigma}_0 = 
{(U_{s}\times \{0\}\times\fg^{s})/ D} \eqno(3.31)$$
characterized by  $\pi_D \circ {\widetilde{\rho}} \circ i_{\mu}=\widehat{\rho}
\circ \pi_{\mu}.$  Hence $X^{\kappa}_{\mu}$ admits a realization in the
Poisson variety $A\Gamma^{\sigma}_0$.

We shall use the following notation.  For $f \in I(\fg)$, the unique
function in $C^{\infty} (A\Gamma^{\sigma}_0)$ determined by 
$i_D^* Pr_3^* {\widetilde{f}}$ (${\widetilde{f}}= f\mid \fg^{s}$)
will be denoted by $\bar f$; while the unique function in
$C^{\infty} (X^{\kappa}_{\mu})$ determined by $i_{\mu}^* L^* {\widetilde{f}}$ 
will be
denoted by ${\Cal F}_{\mu}$.  From the definitions, we have
$${\Cal F}_{\mu} \circ \pi_{\mu} = ({\widehat{\rho}}^{*} \bar f)
  \circ \pi_{\mu} = i_{\mu}^{*}L^{*} {\widetilde{f}}.\eqno(3.32)$$

\proclaim
{Theorem 3.17} If  $\fh^{*}_{s}\neq \{0\}$, then under assumptions A1-A5,
the map ${\widetilde{\rho}}= \rho\mid X^{\kappa}:X^{\kappa}\longrightarrow
  A\Gamma^{\sigma}$ induces a unique Poisson map 
$\widehat{\rho}: X^{\kappa}_{\mu} \longrightarrow A\Gamma^{\sigma}_0$
such that

\noindent (a) functions ${\Cal F}_{\mu}={\widehat{\rho}}^{*}{\bar f}$, 
$f\in I(\fg)$, 
Poisson commute in 
$(X^{\kappa}_{\mu}, \{\,\cdot,\cdot \,\}_{X^{\kappa}_{\mu}})$,
\smallskip
\noindent (b) if $\psi_t$ is the induced flow on 
${\widetilde{\gamma}}^{-1}(0)=U_{s}\times\{0\}\times \fg^{s}$ 
generated by the Hamiltonian $Pr^*_{3} {\widetilde{f}}$, $f \in I(\fg)$, and 
$\phi_t$ is the Hamiltonian flow of ${\Cal F}=L^{*}{\widetilde{f}}$ on 
$X^{\kappa}$, then
under the flow $\phi_{t}$, we have
$$
\aligned
       &{d\over dt} m(\phi_{t}) = {1\over 2}\iota^{*}
          (df(L(\phi_{t}))+ s^{*}(df(L(\phi_{t})))),\\
       & {d\over dt} \tau(\phi_{t}) = 0, \\
       & {d\over dt} L(\phi_{t}) ={1\over 2} [\,L(\phi_{t}), R(m(\phi_{t}))
         (df(L(\phi_{t}))+s^{*}(df(L(\phi_{t})))]\\ 
       &\qquad\quad\quad \,\, + dR(m(\phi_{t}))(\tau(\phi_{t}))
          (df(L(\phi_{t}))+s^{*}(df(L(\phi_{t}))))
\endaligned
$$
where the term involving $dR$ drops out on ${\widetilde{J}}^{-1}(\mu)$.
Moreover, the  reduction $\phi^{red}_t$ of $\phi_{t} \circ i_{\mu}$ on
$X^{\kappa}_{\mu}$ defined by $\phi^{red}_t \circ \pi_{\mu} = \pi_{\mu} \circ
\phi_{t} \circ i_{\mu}$ is a Hamiltonian flow of ${\Cal F}_{\mu} =
{\widehat {\rho}}^{*} {\bar f}$ and ${\widehat {\rho}}\circ \phi^{red}_{t}
(\pi_{\mu} (x)) = \pi_{D} \circ \psi_{t} ({\widetilde{\rho}}(x))$,\,\, 
$x \in {\widetilde{J}}^{-1}(\mu)$.
\endproclaim

\demo
{Proof} (a) Let $f_1$, $f_2\in I(\fg)$, then it is easy to check that
$Pr^{*}_{3}\widetilde{f}_1$,$Pr^{*}_{3}\widetilde{f}_2$ are
extensions of $\pi^{*}_{D}\bar{f}_1$, $\pi^{*}_{D}\bar{f}_2$
with differentials vanishing on the tangent spaces of the $D$-
orbits.  Therefore, if $x\in {\widetilde{J}}^{-1}(\mu))$, we
have $\tau(x)=0$ by assumption A5 and hence
$$\aligned
&\{{\widehat{\rho}}^{*}{\bar f}_1, {\widehat{\rho}}^{*}
         {\bar f}_2 \}_{X^{\kappa}_{\mu}}\circ \pi_{\mu} (x) \\
     =\,& \{{\bar f}_1, {\overline f}_{2} \}_{A\Gamma^{\sigma}_0} \circ 
         \pi_{D}(\widetilde{\rho} (x)) \\
     =\,&\{Pr^{*}_{3}\widetilde{f}_1, Pr^{*}_{3}\widetilde{f}_2 \}_
         {{A\Gamma}^{\sigma}}(\widetilde{\rho} (x)) \\
     =\,&{1\over 4}< [L(x), R(m(x))(df_1+s^{*}(df_1))], df_2+s^{*}(df_2)>
          -(1\leftrightarrow 2)\\
     =\,&{1\over 4}<R(m(x))(df_1+s^{*}(df_1)), ad^{*}_{L(x)}df_2 -
          s^{*}\circ ad^{*}_{L(x)}df_2> -(1\leftrightarrow 2)\\
     =\,& 0.
\endaligned
$$
\smallskip
(b) The equations of motion is a consequence of Proposition 3.11 and
the fact that $\widetilde{\rho}$ is Poisson.  On the other hand,
the assertion on  $\phi^{red}_t$ is a corollary of Theorem 2.16 in
\c{OR} and the relation 
$\widetilde{\rho}\circ\phi_{t} \circ i_{\mu}
= \psi_{t}\circ \widetilde{\rho}\circ i_{\mu}$.
\pf
\enddemo
\medskip

{\bf Case 2. The case of realization in $(\Gamma,\{\cdot,\cdot\}_{R})$}
\medskip

Let ${\Cal P}:X\longrightarrow \Gamma$ be a realization map of $X$
in the coboundary dynamical Poisson groupoid $(\Gamma,\{\cdot,\cdot\}_{R})$.
Recall from  \c{L1} that $\Gamma$ equipped with the action
$${\Cal B}:H\times \Gamma\longrightarrow \Gamma,\,\,
  {\Cal B}_{h}(u,g,v) = (Ad^{*}_{h^{-1}}u, hgh^{-1},Ad^{*}_{h^{-1}}v)
  \eqno(3.33)$$
is a Hamiltonian $H$-space with equivariant momentum map
$$\alpha -\beta:\Gamma\longrightarrow \fh^{*}, (u,g,v)\mapsto u-v.\eqno(3.34)$$

We begin with the following assumption:
\medskip
\noindent G1. there exists a Poisson involution 
$$\Sigma:(\Gamma,\{\cdot,\cdot\}_{R})\longrightarrow
  (\Gamma,\{\cdot,\cdot\}_{R}), (u,g,v)\mapsto
  (s^{*}_{\fh}(v), S(g), s^{*}_{\fh}(u))\eqno(3.35)$$
on $\Gamma$ (where $s$ satisfies the assumptions in Proposition
3.9) such that
$$\Sigma\circ {\Cal P} = {\Cal P}\circ \kappa.\eqno(3.36)$$\newline

Under G1, the map ${\widetilde{\Cal P}}={\Cal P}\mid X^{\kappa}
:X^{\kappa}\longrightarrow \Gamma^{\Sigma}$ is a well-defined Poisson map
by Corollary 3.6,
when the stable loci are equipped with the induced structures.
Let $I(G)$ be the ring of central functions in $G$ and let $I(G^{S})$
consists of restrictions of functions in $I(G)$ to the stable locus
$G^{S}$ of $S$.  If $\bp_{2}$ denote the projection map 
$\Gamma^{\Sigma}\longrightarrow G^{S},\,\,(u,g,s^{*}_{\fh}(u))\mapsto
g$, a natural family of invariant functions on $\Gamma^{\Sigma}$
is $\bp^{*}_{2}I(G^{S})$.  As in the algebroid case, we begin
with a special situation.

\proclaim
{Theorem 3.18} If $s^{*}_{\fh}(u) = u$ for all $u\in U$ so that
$\Gamma^{\Sigma}$ coincides with the gauge group bundle
${\Cal I}\Gamma$ of $\Gamma$, then
the functions in $\bp^{*}_{2}I(G^{S})$ Poisson commutes in
$\Gamma^{\Sigma}\simeq U\times G^{S}$. Therefore, under the
additional assumption that G1 is satisfied,
${\widetilde{\Cal P}}^{*}\bp^{*}_{2}I(G^{S})$
is a Poisson commuting family of functions on $X^{\kappa}.$
\endproclaim

\demo
{Proof} Let $\varphi$, $\psi\in I(G)$ and let 
$\widetilde{\varphi}=\varphi\mid G^S$,
$\widetilde{\psi}= \psi\mid G^S$. Then on using the first
expression in (3.9), we have
$$\aligned
     &\{\bp^{*}_{2}\widetilde{\varphi}, \bp^{*}_{2}\widetilde{\psi}\}_{
        \Gamma^{\Sigma}} (u,g,u)\\
  =\,& <R(u)D'{\widetilde{\bp^{*}_{2}\varphi}},
           D'{\widetilde{\bp^{*}_{2}\psi}}>
     -<R(u)D{\widetilde{\bp^{*}_{2}\varphi}},
           D{\widetilde{\bp^{*}_{2}\psi}}>
\endaligned
$$
where in the second line of the above formula, we have used the
same symbol $\bp_{2}$ to denote the projection map from $\Gamma$
to $G$.  Now, by a direct computation, we can check that
$$D'{\widetilde{\bp^{*}_{2}\varphi}}=D{\widetilde{\bp^{*}_{2}\varphi}}
  ={1\over 2}(D\varphi +s^{*}(D\varphi)).$$
Hence the two terms in the above expression cancel out. The 
second assertion is now clear as ${\widetilde{\Cal P}}$
is Poisson under G1.
\pf
\enddemo

In the general case when the assumption in the above theorem is not
satisfied, we have
$$\eqalign{
     \{\bp^{*}_{2}\widetilde{\varphi}, \bp^{*}_{2}\widetilde{\psi}\}_{
        \Gamma^{\Sigma}} (u,g, s^{*}_{\fh}(u))
  = &\,{1\over 4}<R(s^{*}_{\fh}(u))(D\varphi +s^{*}(D\varphi)),
       D\psi +s^{*}(D\psi)>\cr
    & -{1\over 4}<R(u)(D\varphi +s^{*}(D\varphi)),
           D\psi +s^{*}(D\psi)>\cr}\eqno(3.37)$$
for $\varphi$, $\psi\in I(G)$.  Therefore, 
$\bp^{*}_{2}I(G^{S})$ is no longer a Poisson
commuting family of functions on $\Gamma^{\Sigma}$.  However,
the two terms in (3.37) above do cancel out on 
$\Gamma^{\Sigma}\cap {\Cal I}\Gamma = \{(u,g,u)\mid u\in U_{s},\,g\in G^{S} \}$
where $U_{s}$ is defined in (3.19).  Analogous to Case 1, we now
describe a situation where we can construct Poisson commuting 
functions on a reduced phase space.
To prepare the way for Poisson reduction, we shall  make the following 
assumptions in addition to G1:
\medskip
\noindent G2. the realization map ${\Cal P}$ is $H$-equivariant,\newline
\noindent G3. for some Lie subgroup $D$ of $H$,
$${\Cal B}_{d}\circ \Sigma = \Sigma \circ {\Cal B}_{d},\quad
  {\Cal C}_{d}\circ \kappa = \kappa \circ {\Cal C}_{d},\quad 
  \forall d\in D,\eqno(3.38)$$\newline
\noindent G4. $u-s^{*}_{\fh}(u)\in \frak d^{*}$ for all $u\in U,$
where $\frak d= Lie(D).$

\proclaim
{Proposition 3.19} Under assumptions G1-G4, the stable loci 
$X^{\kappa}$, $\Gamma^{\Sigma}$ are Hamiltonian $D$-spaces
with equivariant momentum maps 
${\widetilde{J}} = i^{*}_{\frak d}\circ (J\mid X^{\kappa})$
and ${\widetilde{\alpha}}-{\widetilde{\beta}} = \alpha - \beta\mid
\Gamma^{\Sigma}$ respectively.  Moreover, the map
$${\widetilde{\Cal P}}={\Cal P}\mid X^{\kappa}
:X^{\kappa}\longrightarrow \Gamma^{\Sigma}\eqno(3.39)$$
is a $D$-equivariant Poisson map.
\endproclaim

\demo
{Proof} The assertion follows from Corollaries 3.6 and 3.8, as in 
Proposition 3.16.
\pf
\enddemo

In order to obtain Poisson commuting functions in the general case, it is 
natural (in view of
the remark after (3.37)) to make the following
additional assumption:
\medskip
\noindent G5. for some regular value $\mu\in\frak d^{*}$ of ${\widetilde{J}},$
$${\widetilde{\Cal P}}({\widetilde{J}}^{-1}(\mu))\subset
 {({\widetilde{\alpha}}-{\widetilde{\beta}})}^{-1}(0)\simeq U_{s}\times G^{S}.
  \eqno(3.40)$$

Analogous to the algebroid case, we have the Poisson variety
$$\left(\Gamma^{\Sigma}_0 = {({\widetilde{\alpha}}-
{\widetilde{\beta}})}^{-1}(0)/D,
  \{\cdot,\cdot\}_{\Gamma^{\Sigma}_0}\right)\eqno(3.41)$$
with the inclusion map $\bi_{D}:{({\widetilde{\alpha}}-
{\widetilde{\beta}})}^{-1}(0)\longrightarrow \Gamma^{\Sigma}$ and
the canonical projection 
$\bpr_{D}:{({\widetilde{\alpha}}-{\widetilde{\beta}})}^{-1}(0)\longrightarrow
\Gamma^{\Sigma}_{0}.$  Moreover, under G5, the map ${\widetilde{\Cal P}}$ 
in Proposition 3.19 induces 
a Poisson map
$${\widehat{\Cal P}}:X^{\kappa}_{\mu}\longrightarrow \Gamma^{\Sigma}_{0}
  \simeq (U_{s}\times G^{S})/D.\eqno(3.42)$$ 

We shall use the following notation.  For $\varphi \in I(G)$, the unique
function in $C^{\infty} (\Gamma^{\Sigma}_0)$ determined by $\bi_D^* 
\bp_2^* {\widetilde{\varphi}}$ (${\widetilde{\varphi}} =\varphi \mid G^{S}$)
will be denoted by $\bar \varphi.$ Also, we set

$$\bL= \bp_2 \circ {\widetilde{\Cal P}} : X^{\kappa} \longrightarrow G^{S}, 
\eqno(3.43)$$
$$ \bm_1 = \widetilde{\alpha} \circ {\widetilde{\Cal P}} : X^{\kappa} 
\longrightarrow U, \eqno(3.44)$$
\noindent and $$ \bm_2 =\widetilde{ \beta} \circ{\widetilde{\Cal P}}  
: X^{\kappa} \longrightarrow U,\eqno(3.45)$$
\noindent i.e. ${\widetilde{\Cal P}} = (\bm_1, \bL, \bm_2)$.
\proclaim
{Theorem 3.20} If $\Gamma^{\Sigma}\neq {\Cal I}\Gamma$, then under
assumptions G1-G5, there exists a unique Poisson structure 
$\{\,\cdot,\cdot \,\}_{X^{\kappa}_{\mu}}$ on the reduced space
$X^{\kappa}_{\mu} = {\widetilde{J}}^{-1}(\mu)/D_{\mu}$
and a unique Poisson map 
${\widehat{\Cal P}}: X^{\kappa}_{\mu} \longrightarrow \Gamma^{\Sigma}_0$
such that

\noindent (a) functions ${\widehat{\Cal P}}^{*}{\bar \varphi}$, $\varphi\in 
I(G)$, 
Poisson commute in 
$(X^{\kappa}_{\mu}, \{\,\cdot,\cdot \,\}_{X^{\kappa}_{\mu}})$,
\smallskip
\noindent (b) if $\psi_{t}$ is the induced flow on 
${({\widetilde{\alpha}}-{\widetilde{\beta}})}^{-1}(0)\subset {\Cal I}\Gamma$
generated by the Hamiltonian $\bp^{*}_{2}\widetilde{\varphi}$, 
$\varphi\in I(G)$ and $\phi_t$ is the Hamiltonian flow of 
$\bL^{*}\widetilde{\varphi}$ on $X^{\kappa}$, then under the flow
$\phi_{t}$, we have
$$\aligned
       &{d\over dt} \bm_1(\phi_{t})={1\over 2}\iota^{*}(D\varphi
         +s^{*}(D\varphi))\\
       &{d\over dt} \bL(\phi_{t}) ={1\over 2}T_{e}l_{\bL(\phi_t)}
         R(s^{*}_{\fh}(\bm_1(\phi_t))(D\varphi +s^{*}(D\varphi))\\
       &\qquad\quad\quad\,\, -{1\over 2}T_{e}r_{\bL(\phi_t)}R(\bm_1(\phi_t))
          (D\varphi +s^{*}(D\varphi))\\
       &{d\over dt} \bm_2(\phi_{t})={1\over 2}\iota^{*}(D\varphi
         +s^{*}(D\varphi)).
\endaligned
$$
\endproclaim

\demo
{Proof} (a) Let $\varphi_1$, $\varphi_2\in I(G)$, then from the
invariance properties of these functions, we can
check that $\bp^{*}_{2}\widetilde{\varphi}_1$,
$\bp^{*}_{2}\widetilde{\varphi}_2$ are extensions of
$\bpr^{*}_{D}\bar{\varphi}_1$, $\bpr^{*}_{D}\bar{\varphi}_2$
with differentials vanishing on the tangent spaces of 
the $D$-orbits.  Since ${\widehat{\Cal P}}$ is Poisson, by making
use of this fact,
it follows that for $x\in {\widetilde{J}}^{-1}(\mu)$, we have
$$\aligned
      & \{{\widehat{\Cal P}}^{*}{\bar{\varphi}}_1,
         {\widehat{\Cal P}}^{*}{\bar{\varphi}}_2\}_{X^{\kappa}_{\mu}}
         \circ \pi_{\mu}(x)\\
   =\, & \{{\bar{\varphi}}_1,{\bar{\varphi}}_2\}_{\Gamma^{\Sigma}_{0}}
         \circ \bpr_{D}({\widetilde{\Cal P}}(x))\\
   =\, & \{\bp^{*}_{2}{\widetilde{\varphi}}_1,
           \bp^{*}_{2}{\widetilde{\varphi}}_2\}_{\Gamma^{\Sigma}}
         \circ {\widetilde{\Cal P}} (x)\\
   =\, & 0
\endaligned
$$
where in the last step we have invoked the formula in (3.37) and
assumption G5.
\newline
(b) This follows from Proposition 3.13 and the fact that 
$\widetilde{\Cal P}$ is Poisson.

\enddemo

\bigskip
\bigskip

\subhead
4. \ Spin Calogero-Moser systems associated with symmetric Lie subal-
\linebreak \phantom{fak} gebras
\endsubhead
\bigskip

A symmetric Lie algebra is a Lie algebra equipped with a Lie algebra
involution.  If $(\fg, \eta)$ is a symmetric Lie algebra, then the
fixed point set $\fg^{\eta}$ will be called a symmetric Lie subalgebra
of $\fg.$  In this section, we shall show that the general scheme in
Section 3 can be applied to  several examples
of spin Calogero-Moser systems associated with real symmetric Lie algebras.  
Because the spin variables of the Dirac reduction belong to 
symmetric Lie subalgebras,  we shall called the reduced
systems  spin Calogero-Moser systems  associated with
symmetric Lie subalgebras.  In the following, we shall restrict
ourselves to the trigonometric case.  It will be clear that the
rational case and the elliptic case can also be handled in a
similar fashion and for this reason, we shall omit the details.
(See Remark 4.1.12(b) and Remark 4.2.7(b) in this connection.)

\subhead
4.1 \ Compact real forms of some spin Calogero-Moser systems
\endsubhead
\bigskip

We begin by introducing a number of Lie-theoretic objects
which we will use throughout the present and the next
subsections.

Let $\fg$ be a complex simple Lie algebra of rank $N$ with Killing form 
$(\cdot,\cdot).$  We fix a Cartan subalgebra  $\fh$ and let 
$\fg=\fh \oplus \sum_{\alpha \in\Delta} \fg_{\alpha}$ be the
root space decomposition of $\fg$ with respect to $\fh.$
For each $\alpha \in \Delta$, denote by $H_{\alpha}$ the element
in $\fh$ which corresponds to $\alpha$ under the isomorphism
between $\fh$ and $\fh^*$ induced by the Killing form $(\cdot,\cdot)$.
On the other hand, for each $\alpha\in \Delta$, we choose root
vectors $e_{\alpha}\in \fg_{\alpha}$ such that for all
$\alpha,\beta \in \Delta$, 
\smallskip
\noindent (i)  $[e_{\alpha}, e_{-\alpha}] =H_{\alpha}$,\newline
\noindent (ii) the constants $N_{\alpha,\beta}$ in the relations
$$[e_{\alpha}, e_{\beta}] = N_{\alpha,\beta}\, e_{\alpha +\beta},
  \quad \alpha, \beta, \alpha+\beta \in \Delta$$
are real and satisfy $N_{\alpha,\beta} =-N_{-\alpha,-\beta}.$

With the notation introduced above, we define
$$\fh_0 =\sum_{\alpha\in\Delta} \Bbb{R} H_{\alpha}.\eqno(4.1.1)$$
Then $(\cdot,\cdot)\mid_{\fh_{0}\times \fh_{0}}$ is positive
definite and each root is real-valued on $\fh_{0}$.  
We shall fix an orthonormal basis $(x_i)_{1\le i\le N}$ of $\fh_{0}$
in what follows.

The Lie algebra $\fg$ has two standard real forms, namely, the normal
real form
$$\fg_{0}= \fh_{0} + \sum_{\alpha\in \Delta} \Bbb{R} e_{\alpha}\eqno(4.1.2)$$
and the compact real form
$$\frak u_{0} = i\fh_{0} + \sum_{\alpha\in \Delta} \Bbb{R} (e_{\alpha}
-e_{-\alpha}) + \sum_{\alpha\in \Delta} \Bbb{R} i(e_{\alpha} +e_{-\alpha}).
 \eqno(4.1.3)$$  
Therefore, if $\fg^{\Bbb{R}}$ denote the algebra $\fg$ regarded as
a real Lie algebra, we have $\fg^{\Bbb{R}}= \fg_{0}\oplus i \fg_{0}
=\frak u_{0}\oplus i \frak u_{0}.$    We shall denote the conjugation
of $\fg$ with respect to $\fg_{0}$ and $\frak u_{0}$ by
$\upsilon$ and $\tau$ respectively.  For simplicity of notation,
we also write
$$\upsilon (q) = \bar q, \quad\,\, \ q\in \fh. \eqno(4.1.4)$$
The pairing on $\fg^{\Bbb R}$ will be taken to be the Killing
form on $\fg^{\Bbb R}$ scaled by the factor $1\over 2$, and will
be denoted by $(\cdot,\cdot)_{\Bbb R}$.  

In addition  to the finite dimensional Lie algebras above, we
will also need their corresponding loop algebras 
$L\fg$, $L\fg^{\Bbb{R}}$ and so on.   For $X\in L\fg$, we shall
write $X(z) = \sum_{- \infty}^{\infty} {X_{n} z^{n}}$ with coefficients
$X_{n}\in \fg$ and similarly for the other loop algebras. Using
the Killing form $(\cdot,\cdot)$ on $\fg$, we can define a
non-degenerate invariant pairing on $L\fg$:
$$(X,Y)_{L\fg} = \sum_{j} (X_{j},Y_{-(j+1)}), \quad X,Y\in L\fg.\eqno(4.1.5)$$ 
Similarly, we have a pairing on $L\fg^{\Bbb{R}}$, given by
$$(X,Y)_{L\fg^{\Bbb{R}}}=\sum_{j} (X_{j},Y_{-(j+1)})_{\Bbb{R}}, \quad X,Y\in 
  L\fg^{\Bbb{R}}.\eqno(4.1.6)$$

In what follows, the connected and simply-connected Lie groups which
integrate the Lie algebras $\fg^{\Bbb{R}}$, $\fh^{\Bbb{R}}$, 
$\fg_{0}$, $\fh_{0}$ and $i\fh_{0}$ will be denoted
by $G^{\Bbb{R}}$, $H^{\Bbb{R}}$, $G_{0}$, $H_{0}$ and $T$ respectively.
On the other hand, $U$ will denote a fixed connected component
of $\{q\in \fh\mid \sin ({1\over 2} \alpha(q)) \neq 0\,\,\,\hbox{for all}
\,\,\,\alpha \in \Delta\}$.  Using
the non-degeneracy of $(\cdot,\cdot)_{\Bbb R}\mid_{\fh^{\Bbb R}\times 
\fh^{\Bbb R}}$,
$(\cdot,\cdot)\mid_{i\fh^{0}\times i\fh^{0}}$
and the pairings above, we shall make the
identifications $(\fg^{\Bbb{R}})^{*}\simeq \fg^{\Bbb{R}}$,
$(\fh^{\Bbb{R}})^{*}\simeq \fh^{\Bbb{R}}$,
$(i\fh_{0})^{*}\simeq i\fh_{0}$,
$\fh^{*}_{0}\simeq \fh_{0}$,
$L{\fg^{\Bbb{R}}}^{*}\simeq L{\fg^{\Bbb{R}}}$, where
$L{\fg^{\Bbb{R}}}^{*}$ is the restricted dual.

Consider the trigonometric dynamical r-matrix with spectral parameter
(which is gauge equivalent to the one in \c{EV}):
$$r(q,z) = \left(c(z)+{1\over 12}z\right)\sum_{i} x_{i}\otimes x_{i} +
 \sum_{\alpha\in \Delta}
  \phi_{\alpha}(q,z)e^{{z\over 12}\alpha(q)} e_{\alpha}\otimes e_{-\alpha} 
  \eqno(4.1.7)$$
where 
$$c(z) = {1\over 2}\cot\left({1\over 2} z\right) \eqno(4.1.8)$$
and
$$
\phi_{\alpha}(q,z)=(c(z)+c(\alpha(q)))\eqno(4.1.9)$$
Then for each $q\in U$, we can define a map 
$r^{\#}_{-}(q):\fg\longrightarrow L\fg$
by the formula
$$((r^{\#}_{-} (q)\xi)(z), \eta) = (r(q,z), \eta \otimes \xi) \eqno(4.1.10)$$
where $\xi$, $\eta \in \fg.$ Therefore, if we write
$\xi = \sum_{i} \xi_{i} x_{i} + \sum_{\alpha\in \Delta} \xi_{\alpha} e_{\alpha}$
for $\xi\in\fg$, we have explicitly that 
$$(r^{\#}_{-} (q)\xi)(z)=d(z)\sum_{i}\xi_{i}x_{i} +\sum_{\alpha\in \Delta}
   \phi_{\alpha}(q,z)e^{{z\over 12}\alpha(q)}\xi_{\alpha}e_{\alpha}
   \eqno(4.1.11)$$
where $d(z) = c(z) + {1\over 12}z$.
We can also construct the associated classical dynamical r-matrix
$R :U\longrightarrow L(L\fg,L\fg)$ for the pair $(L\fg,\fh).$
To do so, we will need to use the following formula:
$$\eqalign{
         &\frac{\partial^{k}r}{\partial z^{k}}(q,z)\cr
      =\,&d^{(k)}(z)\sum_{i}x_{i}\otimes x_{i} +
          \sum_{\alpha\in \Delta}\sum_{j=0}^{k} {k\choose j}
          \phi^{(k-j)}_{\alpha}(q,z)\left({1\over 12}\alpha(q)\right)^{j}
          e^{{z\over 12}\alpha(q)} e_{\alpha}\otimes e_{-\alpha}.\cr} 
  \eqno(4.1.12)$$

\proclaim
{Proposition 4.1.1}  The classical dynamical r-matrix $R$ associated with
the trigonometric dynamical r-matrix with spectral parameter in (4.1.7)
is given by
$$\eqalign{&(R(q)X)(z)\cr =& {1\over 2}X(z) +\sum_{k=0}^{\infty} \frac{
           d^{(k)}(-z)}{k!}\, \Pi_{\fh} X_{-(k+1)}\cr
           & +\sum_{k\geq 0}{1\over k!}\sum_{\alpha \in \Delta}\sum_{j=0}^{k} 
           {k\choose j}\phi^{(k-j)}_{-\alpha}(q,-z)
           \left({1\over 12}\alpha(q)\right)^{j}
            e^{{z\over 12}\alpha(q)}(X_{-(k+1)})_{\alpha} e_{\alpha} .\cr}
           \eqno(4.1.13)$$
\endproclaim

\demo
{Proof} This follows upon substituting the expression for  
$\frac{\partial^{k}r}{\partial z^{k}}(q,z)$ in (4.1.12) into
the formula for $R(q)X$ in Theorem 2.4(b). \pf
\enddemo

Clearly, $R$ induces a map
$U\longrightarrow L(L\fg^{\Bbb{R}},L\fg^{\Bbb{R}})$ which we shall also 
denote by $R$.  

\proclaim
{Proposition 4.1.2} The map
$R:U\longrightarrow L(L\fg^{\Bbb{R}}, L\fg^{\Bbb{R}})$
is a solution of the mDYBE for the pair $(L\fg^{\Bbb{R}}, \fh^{\Bbb{R}})$
with $c= -{1\over 4}$.
Moreover, for $q\in U$, $\xi\in \fg^{\Bbb{R}}$, we have
$r^{\#}_{-}(q)\xi\in L\fg^{\Bbb{R}}.$
\endproclaim

\demo
{Proof} For $q\in U$, $X$, $Y\in L\fg$, the term
$(dR(q)(\cdot)X,Y)_{L\fg}$
which appears in the mDYBE for the pair
$(L\fg,\fh)$ is the unique element in $\fh$
whose pairing with $Z\in \fh$ is given by
$(dR(q)(Z)X,Y)_{L\fg}$.  On the other hand, 
the term $(dR(q)(\cdot)X,Y)_{L\fg^{\Bbb R}}\in \fh^{\Bbb R}$
has a similar meaning.
But now it is easy to show from the non-degeneracy of 
$(\cdot,\cdot)_{\Bbb R}\mid_{\fh^{\Bbb R} \times \fh^{\Bbb R}}$ that
$(dR(q)(\cdot)X,Y)_{L\fg}=(dR(q)(\cdot)X,Y)_{L\fg^{\Bbb R}}.$
Hence it follows from this argument and Theorem 2.4(a) that the map
$R:U\longrightarrow L(L\fg^{\Bbb{R}}, L\fg^{\Bbb{R}})$
is a solution of the mDYBE for the pair $(L\fg^{\Bbb{R}}, \fh^{\Bbb{R}})$.
The assertion involving $r^{\#}_{-}(q)\xi$ is trivial.
\pf
\enddemo

We now introduce the trivial Lie groupoids
$$\Omega = U\times G^{\Bbb{R}}\times U, \quad 
  \Gamma = U\times LG^{\Bbb{R}}\times U.\eqno(4.1.14)$$
By Proposition 4.1.2 and (2.9), we can use the map 
$R:U\longrightarrow L(L\fg^{\Bbb{R}}, L\fg^{\Bbb{R}})$ to construct
the associated coboundary dynamical Lie algebroid
$A^{*}\Gamma\simeq U\times \fh^{\Bbb{R}}\times L\fg^{\Bbb{R}}$.
Hence its dual bundle 
$A\Gamma\simeq U\times \fh^{\Bbb{R}}\times L\fg^{\Bbb{R}}$ has a 
Lie-Poisson structure.  On the other hand, we shall equip the
the dual bundle $A^{*}\Omega\simeq U\times \fh^{\Bbb{R}}\times \fg^{\Bbb{R}}$ 
of the trivial Lie algebroid
$A\Omega$ with the corresponding Lie-Poisson structure.
The Poisson manifolds $A^{*}\Omega$ and $A\Gamma$ are Hamiltonian
$H^{\Bbb{R}}$-spaces with actions
$${\Cal C}_{h} (q,p,\xi) = (q, p, Ad_{h}\xi),\,\, h\in H^{\Bbb{R}},
  (q,p,\xi)\in A^{*}\Omega \eqno(4.1.15)$$
and 
$${\Cal A}_{h} (q,p.X) =(q, p, Ad_{h}X),\,\, h\in H^{\Bbb{R}},
  (q,p,X)\in A\Gamma \eqno(4.1.16)$$ 
and the corresponding equivariant momentum maps are respectively
given by
$$J: A^{*}\Omega\longrightarrow \fh^{\Bbb{R}}, \,\,
  (q,p,\xi)\mapsto \Pi_{\fh^{\Bbb{R}}}\xi\eqno(4.1.17)$$
and
$$\gamma:A\Gamma\longrightarrow \fh^{\Bbb{R}},\,\,
  (q,p,X)\mapsto p \eqno(4.1.18)$$
where  $\Pi_{\fh^{\Bbb{R}}}$ is the projection map to
$\fh^{\Bbb{R}}$ relative to the splitting 
$\fg^{\Bbb R} = \fh^{\Bbb{R}}\oplus (\fh^{\Bbb{R}})^{\perp}$.

In view of our discussion above, the following result is just a real
analog of what we have in Theorem 2.5.

\proclaim
{Proposition 4.1.3} The map
$$\rho: A^{*}\Omega\longrightarrow A\Gamma, \,\,\, (q,-\Pi_{\fh^{\Bbb{R}}}\xi,
  p + r^{\#}_{-}(q)\xi)\eqno(4.1.19)$$
is an $H^{\Bbb{R}}$-equivariant Poisson map. 
\endproclaim

The trigonometric spin Calogero-Moser system which we consider for
our purpose here
is the Hamiltonian system on $A^{*}\Omega$
generated by the Hamiltonian 
$${\Cal H}(q,p,\xi) = Re\left\{{1\over 2} \sum_{i} p^{2}_{i} -{1\over 8} 
   \sum_{\alpha \in \Delta}
  \left(\frac{1}{\sin^{2} {1\over 2}\alpha(q)}-{1\over 3}\right)\xi_{\alpha}
  \xi_{-\alpha}\right\}.\eqno(4.1.20)$$
Let $Q$ be the quadratic
function
$$Q(X)={1\over 2} Re\,\oint_C (X(z), X (z)) \frac{dz}{2\pi i z}, \, \, \,
 X\in L\fg^{\Bbb{R}}\eqno(4.1.21)$$
where $C$ is a small circle around the origin with the positive
orientation.  

The relation between ${\Cal H}$ and $Q$ is given in our next result
which is obtained by a simple residue calculation.

\proclaim
{Proposition 4.1.4} ${\Cal H} =\rho^{*}Pr^{*}_{3}Q$, where 
$Pr_{3}$ is the projection map onto the third factor of
$A\Gamma.$  Thus the Hamiltonian system generated by
${\Cal H}$ can be realized in $A\Gamma.$
\endproclaim

We next examine the phase space underlying ${\Cal H}$.

\proclaim
{Proposition 4.1.5} The map 
$$\kappa : A^{*}\Omega\longrightarrow A^{*}\Omega,\,\,(q,p,\xi)\mapsto
  (\bar q, \bar p, \tau(\xi))\eqno(4.1.22)$$
is a Poisson involution with stable locus
$$A^{*}\Omega^{\kappa} = (U\cap \fh_{0})\times \fh_{0}\times \frak u_{0}.
\eqno(4.1.23)$$
\endproclaim

\demo
{Proof} According to Proposition 3.14, we have to check that
$\tau^{*}$ is a Lie algebra morphism.  But from the orthogonality
of $\frak u_{0}$ and $i\frak u_{0}$ under $(\cdot,\cdot)_{\Bbb R}$,
we can show that $\tau^{*}=\tau$. As $\tau$ is a Cartan involution,
the assertion follows.
\pf
\enddemo

As can be easily verified, the real Hamiltonian system generated
by ${\Cal H}$ has the same equations of motion as the one on 
$U\times \fh\times \fg$ with (complex holomorphic) Hamiltonian
$${\Cal H}^{\Bbb C}(q,p,\xi)={1\over 2} \sum_{i} p^{2}_{i} -{1\over 8} 
   \sum_{\alpha \in \Delta}
  \left(\frac{1}{\sin^{2} {1\over 2}\alpha(q)}-{1\over 3}\right)\xi_{\alpha}
  \xi_{-\alpha}.\eqno(4.1.24)$$
This latter Hamiltonian system, on the other hand, has a {\it compact real 
form\/}, corresponding
to $q_{i}$, $p_{i}\in \Bbb R$, $i=1,\ldots,N$ and $\xi_{\alpha}= 
-\bar{\xi}_{-\alpha}$, $\alpha\in \Delta$.  More precisely, the compact
real form of ${\Cal H}^{\Bbb C}$ is the
Hamiltonian system on $A^{*}\Omega^{\kappa}$ generated by 
$$\eqalign{ \widetilde{\Cal H}(q,p,\xi)
  =\,&({\Cal H}\mid {A^{*}\Omega^{\kappa}})(q,p,\xi)\cr
  =\,&{1\over 2} \sum_{i} p^{2}_{i} +{1\over 8} 
   \sum_{\alpha \in \Delta}
  \left(\frac{1}{\sin^{2} {1\over 2}\alpha(q)}-{1\over 3}\right)
  |\xi_{\alpha}|^{2}.\cr}\eqno(4.1.25)$$
In the rest of the subsection, we shall consider the realization
of this compact real form and its associated integrable model. 
To start with, we introduce the map
$$s: L\fg^{\Bbb{R}}\longrightarrow L\fg^{\Bbb{R}},\,\,
  s(X)(z) = -\tau(X(-\bar z))=-\sum_{j} \tau(X_{j})(-z)^{j}.\eqno(4.1.26)$$

\proclaim
{Proposition 4.1.6} The map
$$\sigma: A\Gamma\longrightarrow A\Gamma, \,\,(q,p,X)\mapsto
  (\bar q, -\bar p, s(X))\eqno(4.1.27)$$
is a Poisson involution with stable locus
$$A\Gamma^{\sigma} = (U\cap \fh_{0})\times i\fh_{0}\times
  (L\fg^{\Bbb{R}})^{s}\eqno(4.1.28)$$
where
$$(L\fg^{\Bbb{R}})^{s} =\lbrace X\in L\fg^{\Bbb{R}}\mid
  X_{2j+1}\in \frak u_{0},\,\, X_{2j}\in i\,\frak u_{0}
 \,\,\,\hbox{for all}\,\, j\,\rbrace.\eqno(4.1.29)$$
Hence $(A^{*}\Gamma, \sigma^{*})$ is a symmetric coboundary
dynamical Lie algebroid.
\endproclaim

\demo
{Proof} It is clear from (4.1.19) that $s$ is an involutive Lie
algebra anti-morphism.  If $Z\in \fh^{\Bbb R}$, we have 
$\tau(Z)= -\overline Z$ from which it follows that 
$s_{\fh^{\Bbb R}}(Z) =\overline Z$. Next, we show that $s^{*}= -s.$
To do so, take $X$, $Y\in L\fg^{\Bbb R}$, then
$$\aligned
         &(s^{*}(X),Y)_{L\fg^{\Bbb R}}\\
     =\, &\sum_{j} (X_j,-\tau(Y_{-(j+1)})(-1)^{j+1})_{\Bbb R}\\
     =\, &\sum_{j} ((-1)^{j}\tau(X_j),Y_{-(j+1)})_{\Bbb R}\\
     =\, &\left(\sum_{j} \tau(X_j)(-z)^{j},Y\right)_{L\fg^{\Bbb R}}\\
     =\, &(-s(X),Y)_{L\fg^{\Bbb R}},
\endaligned
$$
as required. We are now ready to compute $s\circ R(q)\circ s^{*}$ for 
$q\in U$.  Instead of using the explicit expression in (4.1.13), we will
do this using the relationship between $R$ and $r$. This is more
illuminating as the property of $R$ should follow from that of $r$.
To start with, we have
$$\aligned
         & (R(q)s(X))(z)\\
      =\,& {1\over 2}s(X)(z)+\sum_{k\geq 0}\frac{1}{k!}
         \left ( \frac{\partial^{k}r}{\partial z^{k}}(q, -z ), 
         \,(-1)^{k}\tau( X_{-(k+1)})\otimes 1 \right ).
\endaligned
$$   
Therefore, 
$$\aligned
         & (s\circ R(q)\circ s^{*})(X)(z)\\
     =\, & \tau(R(q)s(X)(-\bar z))\\
     =\, & - {1\over 2}X(z) + \sum_{\geq 0}{\frac{1}{k!}}\,\tau
            \left( \frac{\partial^{k}r}{\partial z^{k}}(q, \bar z), 
            \,(-1)^{k}\tau( X_{-(k+1)})\otimes 1 \right ).
\endaligned
$$
To simplify the above expression, note that
$\overline{(a,\tau(\xi))}=(\tau(a),\xi)$
for all $a$,$\xi\in\fg$.  From this relation, we find
$$\tau(a\otimes b,\tau(\xi)\otimes 1) =
(\tau^{\otimes 2}(a\otimes b),\xi\otimes 1)$$
for all $a$,$b$,$\xi\in\fg$.  Consequently, 
$$\aligned
          &\tau\left( \frac{\partial^{k}r}{\partial z^{k}}(q, \bar z), 
            \,(-1)^{k}\tau( X_{-(k+1)})\otimes 1 \right)\\
      =\, &\left(\tau^{\otimes 2}\left( \frac{\partial^{k}r}{\partial z^{k}}
          (q, \bar z)\right), \,(-1)^{k}X_{-(k+1)}\otimes 1 \right).
\endaligned
$$
But from (4.1.12), we can verify that
$$\tau^{\otimes 2}\left(\frac{\partial^{k}r}{\partial z^{k}}(q, \bar z)\right)
  =-(-1)^{k}\frac{\partial^{k}r}{\partial z^{k}}(\bar q, - z).$$
Substitute this into the above expression, we obtain
$$\aligned
         &\tau\left( \frac{\partial^{k}r}{\partial z^{k}}(q, \bar z), 
            \,(-1)^{k}\tau( X_{-(k+1)})\otimes 1 \right)\\
      =\,& - \left(\frac{\partial^{k}r}{\partial z^{k}}
          (\bar q, - z),\,X_{-(k+1)}\otimes 1\right)
\endaligned
$$
and hence that
$$(s\circ R(q)\circ s^{*})(X)(z) = -(R(\bar q)X)(z).$$
Therefore, we can now conclude from Proposition 3.9 that the map $\sigma$ is a
Poisson involution.
\pf
\enddemo

\proclaim
{Proposition 4.1.7} (a) $\sigma\circ \rho = \rho\circ \kappa$.
\smallskip
\noindent (b) For all $d\in T$,
$${\Cal A}_{d}\circ \sigma = \sigma \circ {\Cal A}_{d},\quad
  {\Cal C}_{d}\circ \kappa = \kappa \circ {\Cal C}_{d}.$$
\smallskip
\noindent (c)  $\{Z \in \fh^{\Bbb R}\mid s^{*}_{\fh^{\Bbb R}}(Z) = 
  -Z\} = \frak t.$
\endproclaim

\demo
{Proof} (a) For any $(q,p,\xi)\in A^{*}\Omega$, we have
$$\rho\circ \kappa (q,p,\xi)=\,(\bar q, -\Pi_{\fh^{\Bbb R}} \tau(\xi),
           \bar p + r^{\sharp}_{-}(\bar q)\tau(\xi)).$$
On the other hand,
$$\sigma\circ \rho(q,p,\xi)=\,(\bar q, \overline{\Pi_{\fh^{\Bbb R}}\xi}, 
\bar p + s(r^{\sharp}_{-}(q)\xi)).$$
Now, from the fact that $\tau$ preserves $\fh^{\Bbb R}$, we have
$$\Pi_{\fh^{\Bbb R}}\tau(\xi) = -\overline{\Pi_{\fh^{\Bbb R}}\xi}.$$
Therefore, it remains to show that
$s(r^{\sharp}_{-}(q)\xi) = r^{\sharp}_{-}(\bar q)\tau(\xi).$
To do so, we invoke the explicit expression for $r^{\sharp}_{-}(q)(\xi)$,
according to which 
$$\aligned
         &(r^{\sharp}_{-}(\bar q) \tau(\xi))(z)\\
     =\, &d(z)\sum_{i}\tau(\xi)_{i}x_{i} + 
           \sum_{\alpha\in \Delta} \phi_{\alpha}(\bar q,z)
           e^{{z\over 12}\alpha(\bar q)}(\tau(\xi))_{\alpha}e_{\alpha}\\
     =\, &-d(z)\sum_{i} \bar{\xi}_{i}x_{i} - \sum_{\alpha\in \Delta} 
          \phi_{\alpha}(\bar q,z) e^{{z\over 12}\alpha(\bar q)} 
          {\bar \xi}_{-\alpha}e_{\alpha}.
\endaligned
$$
But on the other hand,
$$\aligned
         &s(r^{\sharp}_{-}(q)\xi)(z)\\
      =\,&-d(-z)\sum_{i}\bar{\xi}_{i}\tau(x_{i}) -\sum_{\alpha\in\Delta} 
           \phi_{\alpha}(\bar q,-z)e^{-{z\over 12}\alpha(\bar q)}
           \bar{\xi}_{\alpha}\tau(e_{\alpha})\\
      =\,&-d(z)\sum_{i}\bar{\xi}_{i}x_{i} - \sum_{\alpha\in \Delta} 
          \phi_{\alpha}(\bar q,z) e^{{z\over 12}\alpha(\bar q)} 
          {\bar \xi}_{-\alpha}e_{\alpha}
\endaligned
$$
and so we have equality.
\newline
(b)  From the definition of $\kappa$ and ${\Cal C}$, it is easy to see
that 
${\Cal C}_{d}\circ \kappa = \kappa \circ {\Cal C}_{d}$\, for
$d\in T$ if and only if $\tau(Ad_d\xi) = Ad_{d}\tau(\xi)$\, for
$d\in T.$  But the latter is clear as $\tau=1$ on $\frak u_{0}$
and $i\fh_{0}\subset \frak u_{0}$. The validity of the other
assertion also boils down to the same condition above, as can be
verified from the definition of $s$.
\newline
(c) From the proof of Proposition 4.1.6, we have 
$s_{\fh^{\Bbb R}}(Z) =\overline Z$ for $Z\in \fh^{\Bbb R}$  from
which the assertion clearly follows.
\pf
\enddemo

Combining Proposition 4.1.3 and Proposition 4.1.7, we conclude that
assumptions A1-A4 are satisfied.  Hence Proposition 3.16 and the
explicit form of $\rho$ give the following result.

\proclaim
{Corollary 4.1.8} The stable loci $A^{*}\Omega^{\kappa}$, 
$A\Gamma^{\sigma}$ are Hamiltonian $T$-spaces with equivariant
momentum maps
${\widetilde{J}}:A^{*}\Omega^{\kappa}\longrightarrow \frak{t},
  \,\,\,(q,p,\xi)\mapsto -\Pi_{\fh^{\Bbb{R}}}\xi,$
and
${\widetilde{\gamma}} = \gamma\mid A\Gamma^{\kappa}$
respectively.  Moreover, the map
$${\widetilde{\rho}}= \rho\mid {A^{*}\Omega^{\kappa}}:
(U\cap \fh_{0})\times \fh_{0}\times \frak u_{0}
  \longrightarrow (U\cap \fh_{0})\times i\fh_{0}\times
  (L\fg^{\Bbb{R}})^{s}\eqno(4.1.30)$$
is a $T$-equivariant Poisson map and 
${\widetilde{\rho}} ({\widetilde{J}}^{-1}(0)) \subset 
 {\widetilde{\gamma}}^{-1}(0).$
\smallskip
\endproclaim

\noindent{\bf Remark 4.1.9.} At this juncture, it is tempting to
reduce $\widetilde{\rho}$ further to the map in (4.2.5) in 
the next subsection.  However, it is not hard to 
show that this cannot be achieved as we cannot find 
appropriate Poisson involutions which commute with $\widetilde{\rho}$. 
\medskip 
We now introduce $L = Pr_{3}\circ {\widetilde{\rho}}$, as in (3.30).  Also,
let $\widetilde Q = Q\mid (L\fg^{\Bbb R})^{s}.$

\proclaim
{Proposition 4.1.10} (a) $\widetilde{\Cal H} = L^{*}\widetilde Q$ and is
invariant under the canonical $T$-action on $A^{*}\Omega^{\kappa}$.
\smallskip
\noindent (b) The restriction of the Hamiltonian equations of motion 
generated by $\widetilde{\Cal H}$ on $A^{*}\Omega^{\kappa}$ to
the invariant submanifold ${\widetilde{J}}^{-1}(0)$ are given by
$$\eqalign{& \dot q = p,\cr
           & \dot p = {1\over 8} \sum_{\alpha \in \Delta}
             \frac{\cot {1\over 2} \alpha(q)}
             {\sin^{2} {1\over 2} \alpha (q)}
             |\xi_{\alpha}|^{2}  H_{\alpha},\cr
           & \dot \xi = \Bigl[\,\xi, -{1\over 4} \sum_{\alpha \in \Delta}
             \frac{\xi_{\alpha}} {\sin^{2} {1\over 2} \alpha (q)}
             e_{\alpha}\,\Bigr].\cr}\eqno(4.1.30)$$
Moreover, under the Hamiltonian flow, we have
$$\dot L (q,p,\xi) = [\, L(q,p,\xi), R(q) M(q,p,\xi) \,]\eqno(4.1.31)$$
on the invariant submanifold  ${\widetilde{J}}^{-1}(0)$ where
$$M(q,p,\xi)(z) = L(q,p,\xi)(z)/z.\eqno(4.1.32)$$
\endproclaim

\demo
{Proof} (a) The assertion is clear from Proposition 4.1.4.
\newline
(b) From the expression for the Poisson bracket in (3.13), the
Hamiltonian equations of motion are given by 
$\dot q = {1\over 2}(\delta_2{\Cal H} + \overline {\delta_2{\Cal H}})$,
$\dot p = -{1\over 2}(\delta_1{\Cal H} + \overline{\delta_1{\Cal H}})$,
and $\dot \xi = [\xi, {1\over 2}(\delta{\Cal H} + \tau(\delta {\Cal H})].$
Therefore, (4.1.30) follows by a direct computation. On the other
hand, (4.1.31) is a consequence of the last equation in Theorem 3.17(b).
\pf
\enddemo

We now restrict ourselves to a smooth component of 
${\widetilde J}^{-1}(0)/T = (U\cap \fh_0)\times \fh_0\times (\frak u_0 \cap
 \fh^{\perp})/T.$  For this purpose, we introduce the following
open submanifold of $\frak u_0$:
$${\Cal U} = \{\,\xi\in \frak u_0 \mid {\xi}_{\alpha_i} = 
(\xi, e_{-\alpha_i})\neq 0, \quad i=1,\ldots, N \,\}. \eqno(4.1.33)$$ 
Clearly, ${\Cal U}$ is dense in $\frak u_0$ and is stable under the
$T$-action.  Therefore, $(U\cap \fh_0)\times \fh_0\times {\Cal U}$
is a Poisson submanifold of $A^{*}\Omega^{\kappa}$, and we can check
that the $T$-action
on  $A^{*}\Omega^{\kappa}$ induces a locally free Hamiltonian group
action on
$(U\cap \fh_0)\times \fh_0\times {\Cal U}.$  Consequently, the corresponding
momentum map is given by the restriction of the one in Corollary 4.1.8.
To simplify notation, we shall denote this momentum map also by
$\widetilde J$ so that 
${\widetilde J}^{-1}(0) = (U\cap \fh_0)\times \fh_0\times 
 ({\Cal U}\cap \fh^{\perp}).$   Now observe that under the $T$-action, all the
isotropy subgroups of the elements of 
${\widetilde J}^{-1}(0)$ are identical.
Since $T$ is compact, it follows from the above observation that
the orbit space
${\widetilde J}^{-1}(0)/T = (U\cap \fh_0)\times \fh_0\times 
 ({\Cal U}\cap \fh^{\perp}/T)$ is a smooth manifold.  We shall
fix a branch of the argument function.  Then the formula
$$ h(\xi) = exp \left(i \sum_{j,k =1}^{N} C_{kj} arg \,\xi_{\alpha_k} 
   h_{\alpha_j}\right)\eqno(4.1.34)$$
defines a map $h : {\Cal U}\cap \fh^{\perp}\longrightarrow T$
where $C = (C_{jk})$ is the inverse of the Cartan matrix and
$h_{\alpha_{j}} = {2 \over (\alpha_j,\alpha_j)} H_{\alpha_{j}}$, 
$j=1,\ldots, N$. Note that for $\xi\in {\Cal U}\cap \fh^{\perp}$,
the element $Ad_{h(\xi)^{-1}} \xi$ is such that
$(Ad_{h(\xi)^{-1}} \xi, e_{-\alpha_i}) > 0, \, i =1,\ldots, N.$  
We shall henceforth identify
the reduced phase space 
${\widetilde J}^{-1}(0)/T = (U\cap \fh_0)\times \fh_0\times 
 ({\Cal U}\cap \fh^{\perp}/T)$
with $(U\cap \fh_0)\times \fh_0\times (\frak u_{0})_{red}$
under the identification map 
$(q,p, [\xi]) \mapsto (q, p, Ad_{h(\xi)^{-1}} \xi)$ 
where 
$$(\frak u_{0})_{red} = \{\,f\in {\Cal U}\cap \fh^{\perp}\mid
  f_{\alpha_i} = (f, e_{-\alpha_i}) > 0, \quad i=1,\ldots, N \,\}.
 \eqno(4.1.35)$$

By Poisson reduction, the reduced manifold 
$(U\cap \fh_0)\times \fh_0\times (\frak u_{0})_{red}$
has a unique Poisson structure where the factor $(\frak u_0)_{red}$
is equipped with the reduction (at 0) of the Lie-Poisson structure
on ${\Cal U}$ by the $T$-action.  Moreover, the reduction of 
the Hamiltonian $\widetilde{\Cal H}$ on $A^{*}\Omega^{\kappa}$
is given by
$$\widetilde{\Cal H}_{0}(q,p,f) = {1\over 2} \sum_{i} p^{2}_{i} +{1\over 8} 
   \sum_{\alpha \in \Delta}
  \left(\frac{1}{\sin^{2} {1\over 2}\alpha(q)}-{1\over 3}\right)
  |f_{\alpha}|^{2}.\eqno(4.1.36)$$

Combining Corollary 4.1.8, Proposition 4.1.10, Theorem 3.17 together with 
our discussion above, we 
therefore obtain the following result.(See the discussion preceding
(3.32) for notations.)

\proclaim
{Theorem 4.1.11} The map $\widetilde{\rho}$ in (4.1.30) restricted
to $(U\cap \fh_0)\times \fh_0\times {\Cal U}$ induces a unique
Poisson map $\widehat{\rho}: (U\cap\fh_0)\times \fh_0\times (\frak u_{0})_{red}
\longrightarrow A\Gamma^{\sigma}_0$ such that

\noindent(a)  functions 
${\Cal F}_0={\widehat{\rho}}^{*}{\bar f}$, $f\in I(L\fg)$, 
Poisson commute in  $(U\cap\fh_0)\times \fh_0\times (\frak u_{0})_{red}$
and provide a family of conserved quantities in involution for the
Hamiltonian $\widetilde{\Cal H}_{0}$,
\smallskip
\noindent (b) Under the Hamiltonian flow generated by
$\widetilde{\Cal H}_{0}$ on $(U\cap \fh_0)\times \fh_0\times (\frak u_{0})_{red}$,
$$\dot L(q,p,f) = [\,L(q,p,f), R(q)M(q,p,f) + {\Cal M}\,]$$
where
$${\Cal M} ={i\over 4} \sum_{j,k} \frac{C_{kj}}{f_{\alpha_{k}}}
          \sum \Sb \alpha \in \Delta \\
           \alpha_{k}-\alpha \in \Delta \endSb N_{\alpha, \alpha_{j}-\alpha} 
           \frac{Im(f_{\alpha} f_{\alpha_{k}-\alpha})} 
           {\sin^{2}{1\over 2} \alpha (q)}  
           h_{\alpha_j}.$$ 
\endproclaim

\demo
{Proof} Only (b) requires a proof.  First of all, from (4.1.31) and the
$T$-equivariance of $\widetilde{\rho}$ and $R$, we have (for
$f=Ad_{h(\xi)^{-1}}\xi$):
$$\aligned
       &\dot L(q,p,f)\\
   =\, &\Bigl[\,L(q,p,f), R(q)M(q,p,f) + T_{h(\xi)} l_{h(\xi)^{-1}}{d\over dt}
       h(\xi)\,\Bigr].
\endaligned$$
Now, differentiating $h(\xi)$  yields
$$T_{h(\xi)}l_{h(\xi)^{-1}} {d\over dt}h(\xi) = i\sum_{j,k} C_{kj} 
(arg\, \xi_{\alpha_{k}})\spdot \, h_{\alpha_{j}}.\quad\quad\quad\quad (*)$$
But
$$\aligned
\dot \xi_{\alpha_k}  &= \Bigl(\,\Bigl[\,\xi,  -{1\over 4} 
                     \sum_{\alpha \in \Delta}
                   \frac{\xi_{\alpha}} {\sin^{2} {1\over 2} \alpha (q)}
                    e_{\alpha}\,\Bigr], e_{-\alpha_k}\,\Bigr)\\
                   &= {1\over 4} 
                    \sum \Sb \alpha \in \Delta \\
                   \alpha_{k}-\alpha \in \Delta \endSb 
                   N_{\alpha, \alpha_{k}-\alpha} 
                   \frac{\xi_{\alpha} \xi_{\alpha_{k}-\alpha}} 
                    {\sin^{2}{1\over 2} \alpha (q)}.
\endaligned
$$ 
Therefore, upon dividing both sides of the above expression by
$e^{i arg \,\xi_{\alpha_{k}}}$ and taking the imaginary part of both sides,
we find
$$f_{\alpha_k}(arg\,\xi_{\alpha_{k}})\spdot  ={1\over 4} 
                    \sum \Sb \alpha \in \Delta \\
                   \alpha_{k}-\alpha \in \Delta \endSb 
                   N_{\alpha, \alpha_{k}-\alpha} 
                   \frac{Im(f_{\alpha} f_{\alpha_{k}-\alpha})} 
                    {\sin^{2}{1\over 2} \alpha (q)}$$
where we have used the reality of $N_{\alpha,\beta}$ and $\alpha(q)$
together with the fact that 
$f = \sum_{\alpha\in \Delta} \xi_{\alpha} e^{-i\sum_{k} m^{k}_{\alpha} 
arg \,\xi_{\alpha_{k}}} 
 \,e_{\alpha}$  (the $m^{k}_{\alpha}$ are defined by 
$\alpha = \sum_{k} m^{k}_{\alpha} e_{\alpha_{k}}$).  Consequently,
when we substitute this in (*), the desired expression for
${\Cal M}$ follows.
\pf
\enddemo

\noindent{\bf Remark 4.1.12.} (a) The Hamiltonian $\widetilde{\Cal H}_{0}$
is in fact completely integrable in the sense of Liouville on
generic symplectic leaves of the reduced phase space.  The
same remark also applies to the integrable spin Calogero-Moser systems
in \c{LX2} for all simple Lie algebras.  A unifying and representation
independent method to establish the Liouville integrability of such
systems for all simple Lie algebras will be given in a forthcoming
paper. For  $\fg = gl(N,\Bbb{C})$ with $\fh$ taken to be the set
of diagonal matrices in $\fg$, a sketch of the proof will be
given below.
\smallskip
\noindent (b) For the rational dynamical r-matrix
${\Omega \over z} + \sum_{\alpha\in \Delta}
  {1 \over \alpha(q)} e_{\alpha} \otimes e_{-\alpha}$ and
the elliptic dynamical r-matrix
$\zeta(z)\sum_{i} x_{i}\otimes x_{i} -\sum_{\alpha\in \Delta}
  l(\alpha(q),z) e_{\alpha}\otimes e_{-\alpha}$
(here $l(w,z) = -\frac{\sigma(w+z)}{\sigma(w)\sigma(z)}$), 
recall that we
can associate the corresponding (complex holomorphic) spin
Calogero-Moser systems \c{LX2}.  We remark that the compact real
forms of these Hamiltonian systems can also be treated in the
same way. Indeed, with the corresponding $r^{\sharp}_{-}(q)$ and
$R(q)$, our analysis above can be repeated and everything goes
through just the same as before.  Note that 
the explicit form of $r(q,z)$ is only used in checking
$$\tau^{\otimes 2}\left(\frac{\partial^{k}r}{\partial z^{k}}(q, \bar z)\right)
  =-(-1)^{k}\frac{\partial^{k}r}{\partial z^{k}}(\bar q, - z)$$
and in verifying 
$s(r^{\sharp}_{-}(q)\xi) = r^{\sharp}_{-}(\bar q)\tau(\xi).$  Finally we
remark that a version of the rational spin Calogero-Moser system 
similar to the compact real form of our rational case has
been obtained in \c{AKLM} by reducing a free Hamiltonian system
on a cotangent bundle.  However, it is not clear how this method
can be generalized to handle the elliptic case.  
\medskip
In the remainder of the subsection, we shall give a brief sketch of
the Liouville integrability for the reductive case where $\fg =gl(N,\Bbb{C}).$
Indeed, it is easy to see that we can repeat the same analysis above for 
this case with $\fh$ taken to be the set of
all diagonal matrices in $\fg$ and with the trigonometric dynamical
r-matrix with spectral parameter 
$$r(q,z) = \left(c(z)+{1\over 12}z\right)\sum_{i} e_{ii}\otimes e_{ii} +
 \sum_{i\neq j}(c(z)+c(q_i-q_j))e^{{z\over 12}(q_i-q_j)} e_{ij}\otimes e_{ji}$$
where $e_{ij}$ is the $N\times N$ matrix with a $1$ in the $(i,j)$ entry
and zeros elsewhere and $c(z)$ is as in (4.1.8).  In this case, $\fh_0$
and $\frak u_0$ are the subalgebras of $\fg^{\Bbb{R}}$ consisting of
real diagonal matrices and skew-Hermitian matrices respectively and we
take $U$ to be a fixed connected component of 
$\{q\in \fh\mid \sin (\frac{q_i-q_j}{2}) \neq 0\,\,\,\hbox{for all}
\,\,\, i\neq j\}$.  Moreover,  $\fg_0 = gl(N,\Bbb{R})$  and
$T$ is the maximal torus of the unitary group $U(N)$ consisting of unitary
diagonal matrices.  For our analysis below, we also have to introduce the
torus $T^{\prime}\subset T$ consisting of matrices of the form
$diag(1,e^{i\theta_2},\ldots,e^{i\theta_N})$.
Clearly, the above results for simple Lie algebras have obvious analogs in
this case.  In particular, we can obtain the Hamiltonian
$$ \widetilde{\Cal H}(q,p,\xi)
  =\,{1\over 2} \sum_{i} p^{2}_{i} +{1\over 8} 
   \sum_{i\neq j}
  \left(\frac{1}{\sin^{2} (\frac{q_i-q_j}{2})}-{1\over 3}\right)
  |\xi_{ij}|^{2}$$
on $A^{*}\Omega^{\kappa} =(U\cap\fh_0)\times \fh_0\times\frak u_0$
and its associated realization map 
$$\widetilde{\rho}:(U\cap \fh_{0})\times \fh_{0}\times \frak u_{0}
  \longrightarrow (U\cap \fh_{0})\times i\fh_{0}\times
  (L\fg^{\Bbb{R}})^{s}$$
by Dirac reduction.  For the Hamiltonian system generated by
$\widetilde{\Cal H}$, we note that
the restriction of its  equations of motion to the
invariant submanifold $\widetilde{J}^{-1}(0)$ are given by
$$\eqalign{& \dot q = p,\cr
           & \dot p = {1\over 4} \sum_{i\neq j}
             \frac{\cot (\frac{q_i-q_j}{2})}
             {\sin^{2} (\frac{q_i-q_j}{2})}
             |\xi_{ij}|^{2}e_{ii},\cr
           & \dot \xi = \Bigl[\,\xi, -{1\over 4} \sum_{i\neq j}
             \frac{\xi_{ij}} {\sin^{2}(\frac {q_i-q_j}{2})}
             e_{ij}\,\Bigr].\cr}\eqno(4.1.37)$$
Thus the equations coincide exactly with the ones derived in \c{H} (cf. also
\c{HKS} and \c{NM}) for
the eigenphases  and  (essentially) the eigenvectors of
the unitary Floquet operator $F = e^{-i\lambda V}e^{-iH_0}$
(as a function of $\lambda$) associated with a periodically kicked 
quantum system if we take the time variable in (4.1.37) to be the kick 
strength $\lambda.$  More importantly, we have the Lax operator
$L = Pr_{3}\circ \widetilde{\rho}$ whose restriction to 
$\widetilde{J}^{-1}(0) = (U\cap \fh_0)\times \fh_0\times (\frak u_0 \cap 
 \fh^{\perp})$ is given explicitly by
$$L(q,p,\xi)(z) = p + \sum_{i\neq j}
   (c(z) + c(q_i-q_j))e^{{z\over 12}(q_i-q_j)}\xi_{ij}e_{ij}\eqno(4.1.38)$$
and we can establish Liouville integrability for the
reduced Hamiltonian system using $L$.
To do so, we introduce the open submanifold of $\frak u_0$: 
$${\Cal U} = \{\,\xi\in \frak u_0 \mid {\xi}_{i,i+1}  \neq 0, 
   \quad i=1,\ldots, N \,\}. \eqno(4.1.39)$$ 
Then ${\Cal U}$ is dense in $\frak u_0$ and is stable under the
$T$-action.  Hence $(U\cap \fh_0)\times \fh_0\times {\Cal U}$ is
a Poisson submanifold of $A^{*}\Omega^{\kappa}$ and the $T$-action
on $A^{*}\Omega^{\kappa}$
induces a Hamiltonian group action on 
$(U\cap \fh_0)\times \fh_0\times {\Cal U}.$  Denote the momentum
map of this action also by $\widetilde{J}$.  Then
$\widetilde{J}^{-1}(0) = (U\cap \fh_0)\times \fh_0\times ({\Cal U}\cap 
\fh^{\perp}).$  Since the isotropy subgroups of the elements of
$\widetilde{J}$ under the $T$-action are all equal to
the center of $U(N)$ ($=\{e^{i\theta}I\}\simeq S^{1}$) and $T$ is
compact, we conclude that the orbit
space $\widetilde{J}^{-1}(0)/T =(U\cap \fh_0)\times \fh_0\times ({\Cal U}\cap 
\fh^{\perp}/T)$  is a smooth manifold. Indeed, 
$\widetilde{J}^{-1}(0)/T = \widetilde{J}^{-1}(0)/T^{\prime}$ and
we can check that the action of $T^{\prime}$ on $\widetilde{J}^{-1}(0)$ is free.
Let 
$$(\frak u_0)_{red} = \left\{f \in{\Cal U}\cap \fh^{\perp}\mid
   f_{i,i+1} > 0,\quad i=1,\ldots, N-1\right\}.\eqno(4.1.40)$$
Note that for 
$\xi\in {\Cal U}\cap \fh^{\perp}$, there exists unique $h(\xi)\in T^{\prime}$
such that $Ad_{h(\xi)^{-1}}\xi \in (\frak u_0)_{red}.$  We shall
henceforth identify  $\widetilde{J}^{-1}(0)/T$ with
$(U\cap \fh_0)\times \fh_0\times (\frak u_0)_{red}.$

Now the dimension of the generic symplectic leaves of  
$(U\cap \fh_0)\times \fh_0\times (\frak u_0)_{red}$
is  given by  
$$2N + dim_{\Bbb{R}}\, \frak u_0 - N - 2(N-1)= N(N-1) +2.\eqno(4.1.41)$$
Therefore, in order to show that the reduced Hamiltonian 
$\widetilde{\Cal H}_0$
is completely integrable in the sense of Liouville on these
generic symplectic leaves,  we have
to exhibit $1 + {N(N-1)\over 2}$ functionally independent conserved
quantities in involution.  To do so, we define $\widetilde{L}$
by $\widetilde{L}(q,p,f)(z) =  Ad_{e^{-{z\over 12}q}} L(q,,p,f)(z)$
for $(q,p,f)\in(U\cap \fh_0)\times \fh_0\times (\frak u_0)_{red}.$ 
Then the characteristic polynomial of $\widetilde{L}(q,p,f)(z)$
has the form
$$det(\widetilde{L}(q,p,f)(z) -w) =\sum_{r=0}^{N}\sum_{k=0}^{r} 
I_{rk}(q,p,f)\, c(z)^{k} w^{N-r}.\eqno(4.1.42)$$
Now observe that 
$(\widetilde{L}(q,p,f)(z))^{*} = \widetilde{L}(q,p,f)(-\bar z)$.
Hence the functions $I_{r,2k}(q,p,f)$, $i I_{r,2k+1}(q,p,f)$ are 
real valued  and provide the conserved quantities in
involution.  Clearly, among these are the $N-1$ Casimirs
$I_{2k,2k}$ and $i I_{2k+1,2k+1}$ ,$k\geq 1$ (note that $I_{11} =0$ on
$(U\cap \fh_0)\times \fh_0\times (\frak u_0)_{red}$).   On the other hand, 
it follows
from the explicit expression of $\widetilde{L}$ that
$$Ad_{e^{iq}}\widetilde{L}(i\infty) = \widetilde{L}(-i\infty).\eqno(4.1.43)$$
Consequently, we obtain the relations
$$\sum_{k=0}^{\left[r-1\over 2\right]} (-1)^{k} I_{r,2k+1}(q,p,f) = 0\eqno(4.1.44)$$
for $r = 1,\ldots,N$.  Hence the total number of independent
nontrivial integrals equals
$$1 + \sum_{r=2}^{N} (r-1) = 1 + {1\over 2} N(N-1),\eqno(4.1.45)$$
as required.

\bigskip
\bigskip

\subhead
4.2 \ Normal compact forms of some spin Calogero-Moser systems
\endsubhead
\bigskip

It is clear from (4.1.13) that $R$ also induces a map
$U\cap\fh_{0}\longrightarrow L(L\fg_{0},L\fg_{0})$
which we will also denote by $R$.

\proclaim
{Proposition 4.2.1} The map 
$R:U\cap \fh_{0}\longrightarrow L(L\fg_{0},L\fg_{0})$ is a solution
of the mDYBE for the pair $(L\fg_{0},\fh_{0})$ with $c=-{1\over 4}.$  Moreover, 
for $q\in U\cap\fh_{0}$, $\xi\in \fg_{0}$, we have
$r^{\#}_{-}(q)\xi\in L\fg_{0}.$
\endproclaim

\demo
{Proof} For $q\in U\cap \fh_{0}$, $X$, $Y\in L\fg_{0}$, the element
$(dR(q)(\cdot)X,Y)$ must lie in $\fh_{0}$ because for 
$Z\in \fh_{0}$, $(dR(q)(Z)X,Y)\in \Bbb{R}$. The other
assertion is clear from (4.1.11) as $\alpha(q)\in\Bbb{R}$ for
$q\in U\cap\fh_{0}$.
\pf
\enddemo 

We next introduce the trivial Lie groupoids
$$\Omega = (U\cap\fh_0)\times G_{0}\times (U\cap\fh_0), \quad 
  \Gamma = (U\cap\fh_0)\times LG_{0}\times (U\cap\fh_0).\eqno(4.2.1)$$
By Proposition 4.2.1 and (2.9), we can equip the dual bundle 
$A^{*}\Gamma\simeq (U\cap\fh_{0})\times \fh_{0}\times L\fg_{0}$
of $A\Gamma$ with the Lie algebroid structure associated to 
$R:U\cap \fh_{0}\longrightarrow L(L\fg_{0},L\fg_{0})$.  Hence its dual bundle 
$A\Gamma\simeq (U\cap \fh_0)\times \fh_{0}\times L\fg_{0}$ has a 
Lie-Poisson structure.   We shall also equip the
the dual bundle 
$A^{*}\Omega\simeq (U\cap\fh_{0})\times \fh_{0}\times \fg_{0}$ 
of the trivial Lie algebroid
$A\Omega$ with the corresponding Lie-Poisson structure.
The Poisson manifolds $A^{*}\Omega$ and $A\Gamma$ are Hamiltonian
$H_{0}$-spaces.  Indeed, the actions are defined by expressions
identical to (4.1.15) and (4.1.16) provided that we change
$H^{\Bbb{R}}$ to $H_{0}$ and use the definitions of 
$\Omega$ and $\Gamma$ in (4.2.1).  On the other hand,
the corresponding equivariant momentum maps are given by
$$J: A^{*}\Omega\longrightarrow \fh_{0}, \,\,
  (q,p,\xi)\mapsto \Pi_{\fh_{0}}\xi\eqno(4.2.2)$$
and
$$\gamma:A\Gamma\longrightarrow \fh_{0},\,\,
  (q,p,X)\mapsto p \eqno(4.2.3)$$
where $\Pi_{\fh_{0}}$ is the projection map to $\fh_{0}$ relative
to the decomposition $\fg_{0} = \fh_{0}\oplus (\fh_{0})^{\perp}$.

As in Propositions 4.1.3 and 4.1.4, we have the following result in
this case.

\proclaim
{Proposition 4.2.2} The map
$$\rho: A^{*}\Omega\longrightarrow A\Gamma, \,\,\, (q,-\Pi_{\fh_{0}}\xi,
  p + r^{\#}_{-}(q)\xi)\eqno(4.2.4)$$
is an $H_{0}$-equivariant Poisson map.  Moreover, the map
$\rho$ gives a realization of the spin Calogero-Moser system
on $A^{*}\Omega$ with Hamiltonian
$${\Cal H}(q,p,\xi) = {1\over 2} \sum_{i} p^{2}_{i} -{1\over 8} 
  \sum_{\alpha \in \Delta}
  \left(\frac{1}{\sin^{2} {1\over 2}\alpha(q)}-{1\over 3}\right)\xi_{\alpha}
  \xi_{-\alpha}\eqno(4.2.5)$$
in $A\Gamma.$
\endproclaim

The Hamiltonian system in (4.2.5) will be called the {\it normal real
form\/} of the complex holomorphic system ${\Cal H}^{\Bbb C}$ in
(4.1.24).

In the rest of the section, we shall reduce this normal real form
to what we call the {\it normal compact form\/}.  As the
reader will see, the normal compact form has a natural family
of conserved quantities in involution.

For this purpose, we introduce 
$$\frak{k}_{0}=\sum_{\alpha\in \Delta} \Bbb{R}(e_{\alpha}-e_{-\alpha}),
  \,\, \frak{p}_{0}=\fh_{0} + \sum_{\alpha\in \Delta}\Bbb{R}(e_{\alpha}
  +e_{-\alpha}).\eqno(4.2.6)$$
Then
$$\fg_{0}=\frak{k}_{0}+\frak{p}_{0}\eqno(4.2.7)$$
is a Cartan decomposition of $\fg_{0}$.  Let $\theta$ be the corresponding
involution.

\proclaim
{Proposition 4.2.3} The map 
$$\kappa : A^{*}\Omega\longrightarrow A^{*}\Omega,\,\,(q,p,\xi)\mapsto
  ( q, p, \theta(\xi))\eqno(4.2.8)$$
is a Poisson involution with stable locus
$$A^{*}\Omega^{\kappa} = (U\cap \fh_{0})\times \fh_{0}\times \frak k_{0}.
\eqno(4.2.9)$$
\endproclaim

\demo
{Proof} Since $\theta$ is a Cartan involution, it follows that
$\frak k_{0}$ and $\frak p_{0}$ are orthogonal under 
$(\cdot,\cdot)\mid_{\fg_0\times \fg_0}$. Using this property, we can
show that $\theta^{*}=\theta.$  Consequently, we conclude from
Proposition 3.14 that $\kappa$ is a Poisson involution.
\pf
\enddemo

As $\frak k_{0} = \frak u_{0} \cap \fg_{0}$, we shall call $\frak k_{0}$
the {\it normal compact form\/} of $\fg.$  Note, however, that
$\frak k_{0}$ is not at all a real form of $\fg$ because its
complexification is different from $\fg.$  In view of this
terminology, we define the
{\it normal compact form\/} of ${\Cal H}^{\Bbb C}$ to be the Hamiltonian system
on $A^{*}\Omega^{\kappa}$ generated by 
$$\eqalign{ \widetilde{\Cal H}(q,p,\xi)
  =\,&({\Cal H}\mid {A^{*}\Omega^{\kappa}})(q,p,\xi)\cr
  =\,&{1\over 2} \sum_{i} p^{2}_{i} +{1\over 8} 
   \sum_{\alpha \in \Delta}
  \left(\frac{1}{\sin^{2} {1\over 2}\alpha(q)}-{1\over 3}\right)
  |\xi_{\alpha}|^{2}.\cr}\eqno(4.2.10)$$

In order to discuss the realization of this Hamiltonian system,
we introduce
$$s: L\fg_{0}\longrightarrow L\fg_{0},\,\,
  s(X)(z) = -\sum_{j} \theta(X_{j})(-z)^{j}.\eqno(4.2.11)$$

\proclaim
{Proposition 4.2.4} The map
$$\sigma: A\Gamma\longrightarrow A\Gamma, \,\,(q,p,X)\mapsto
  ( q, - p, s(X))\eqno(4.2.12)$$
is a Poisson involution with stable locus
$$A\Gamma^{\sigma} = (U\cap \fh_{0})\times \{0\}\times
  (L\fg_{0})^{s}\eqno(4.2.13)$$
where
$$(L\fg_{0})^{s} =\lbrace X\in L\fg_{0}\mid
  X_{2j+1}\in \frak k_{0},\,\, X_{2j}\in \frak p_{0}
 \,\,\,\hbox{for all}\,\, j\,\rbrace.\eqno(4.2.14)$$
Consequently, $Pr_{3}^{*} I((L\fg_{0})^{s})$ is a Poisson commuting
family of functions on $A\Gamma^{\sigma}$.
\endproclaim

\demo
{Proof} Since $\theta=-1$ on $\frak p_{0}$ and $\fh_{0}\subset \frak p_{0}$,
we have $s_{\fh_0} = -\theta\mid_{\fh_0} = id_{\fh_0}.$
Therefore, $s^{*}_{\fh_0}=id_{\fh_0}$.  The rest of the proof
of the first assertion is similar to the one for Proposition 4.1.6.
On the other hand, the second assertion is just a consequence of
(4.2.13) and Theorem 3.15.
\pf
\enddemo

Our next result shows that assumption A1 is satisfied.  Its proof is similar 
to the one for Proposition 4.1.7.

\proclaim
{Proposition 4.2.5} $\sigma\circ \rho =\rho\circ \kappa$.
\endproclaim

From this proposition and (4.2.13), we see that the assumptions in
Theorem 3.15 are satisfied.  Hence we obtain the following result.

\proclaim
{Theorem 4.2.6} (a) The map
$${\widetilde{\rho}}= \rho\mid {A^{*}\Omega^{\kappa}}:
(U\cap \fh_{0})\times \fh_{0}\times \frak k_{0}
  \longrightarrow (U\cap \fh_{0})\times \{0\}\times
  (L\fg_{0})^{s}\eqno(4.2.15)$$
is a  Poisson map.
\smallskip
\noindent (b) $\widetilde{\Cal H} = L^{*}\widetilde Q$ and 
admits $L^{*}I((L\fg_{0})^{s})$ as a family of conserved quantities
in involution.  Here, $L = Pr_{3}\circ {\widetilde{\rho}}$ and
 $\widetilde Q = Q\mid (L\fg_{0})^{s}.$
\endproclaim

\noindent{\bf Remark 4.2.7.} (a) The Hamiltonian system generated
by $\widetilde{\Cal H}$ is completely integrable in the sense
of Liouville on generic symplectic leaves of $A^{*}\Omega^{\kappa}$.
This will also be treated in the forthcoming paper which we
mentioned in the previous subsection.
\smallskip
\noindent (b) The normal compact forms of the rational and the
elliptic spin calogero-Moser systems (corresponding to
the dynamical r-matrices in Remark 4.1.12(b)) can also treated
in a similar fashion.  Note that for the elliptic case, in
order for the corresponding $R:U\longrightarrow L(L\fg, L\fg)$
to induce a map 
$U\cap\fh_{0}\longrightarrow L(L\fg_{0},L\fg_{0}),$ we have to
make an additional assumption, namely, we have to restrict to
periods $2\omega_1$, $2\omega_2$ (of the elliptic functions)
for which the invariants $g_2 = 60\sum_{\omega\in \Lambda\backslash\{0\}} \omega^{-4}$ 
and $g_3 = 140\sum_{\omega\in \Lambda\backslash\{0\}} \omega^{-6}$ are real,
where $\Lambda = 2\omega_{1}\Bbb{Z} + 2\omega_{2}\Bbb{Z}$.  
\medskip
As in Section 4.1, we shall close this subsection with a sketch of 
the Liouville integrability for  $\fg = gl(N, \Bbb{C})$. 
In this case,  $U$, $\fh_0$ and $r(q,z)$ are the same objects which 
appear at the end of section 4.1 and we have $\fg_0 = gl(N, \Bbb{R})$. 
Thus the factors $\frak{k}_{0}$ and $\frak{p}_0$ in the Cartan decomposition
are respectively the set of skew-symmetric matrices and symmetric
matrices in $\fg_0$.  Clearly the results above for simple Lie algebras
have obvious analogs
in this case.  In particular, for the Hamiltonian system on
$A^{*}\Omega^{\kappa} = (U\cap \fh_0)\times \fh_0\times \frak{k}_0$
generated by 
$$ \widetilde{\Cal H}(q,p,\xi)
  =\,{1\over 2} \sum_{i} p^{2}_{i} +{1\over 8} 
   \sum_{i\neq j}
  \left(\frac{1}{\sin^{2} (\frac{q_i-q_j}{2})}-{1\over 3}\right)
  |\xi_{ij}|^{2},\eqno(4.2.16)$$
its Hamiltonian equations of motion are given by the same expressions in
(4.1.37) but with $\xi\in \frak{k}_0$.  In this case, the equations are 
associated
with an orthogonal Floquet operator $F = e^{-i\lambda V}e^{-iH_0}$
and the Lax operator $L = Pr_{3}\circ \widetilde{\rho}$ on
$A^{*}\Omega^{\kappa}$ has the same form as the one in 
(4.1.38) but with $\xi\in \frak{k}_0$. Now the dimension of the generic 
symplectic leaves
of  $A^{*}\Omega^{\kappa}$ is given by
$2N + \frac{N(N-1)}{2} - \left[{N\over 2}\right] = 2N + 2\left[(N-1)^{2}\over 4
 \right].$  In order to show $\widetilde{\Cal H}$ is completely
integrable in the sense of Liouville on these leaves, we have to
exhibit $N + \left[(N-1)^{2}\over 4
 \right]$ functionally independent conserved quantities in involution.
Put $\widetilde{L}(q,p,\xi)(z) = Ad_{e^{-{z\over 12}q}} L(q,,p,\xi)(z),$
then it is easy to check that
 $(\widetilde{L}(q,p,\xi)(z))^{T} = \widetilde{L}(q,p,\xi)(-z)$.
Therefore the characteristic polynomial of $L(q,p,\xi)(z)$ is an
even function of $z$.
Hence we have
$$det(\widetilde{L}(q,p,\xi)(z) -w) =\sum_{r=0}^{N}\sum_{k=0}^{\left[
 r\over 2\right]} I_{rk}(q,p,\xi)\, c(z)^{2k} w^{N-r}\eqno(4.2.17)$$
and the $I_{rk}$'s are conserved quantities in involution.
Clearly, the functions
$I_{2k,k},$$k=1,\ldots,\left[N\over 2\right]$, 
are Casimirs.  Therefore, the total number of nontrivial integrals
is given by
$$\sum_{r=1}^{N} \left(\left[r\over 2\right] + 1\right) -\left[N\over 2\right]
  = N + \left[(N-1)^{2}\over 4\right].\eqno(4.2.18)$$
\bigskip
\bigskip

\subhead
5. \ Symmetric space  spin Ruijsenaars-Schneider models and soliton 
dy- \linebreak \phantom{fak}\,\, namics of affine Toda field theory
\endsubhead
\bigskip

There is a well-known correspondence between the $N$-soliton solutions 
of the $A^{(1)}_n$ affine Toda field theory and some spin-generalized 
Ruijensaars-Schneider equations \c{BH}.  The goal of this section is
to resolve a long-standing problem regarding the Hamiltonian formulation
and the integrability of such equations.

Let $\fg = gl(N,\Bbb{C})$, and let $\fh$ be the Cartan subalgebra
of $\fg$ consisting of diagonal matrices.  We shall denote by
$\fg^{\Bbb{R}}$ (resp. $\fh^{\Bbb{R}}$) the algebra $\fg$ 
(resp. $\fh$) regarded as a real Lie
algebra.  It is well-known that
$$\frak u(N) = \lbrace \,N\times N \,\,\hbox{skew-Hermitian matrices\,} \rbrace
  \eqno(5.1)$$ 
is a compact real form of $\fg$.  We shall denote by $\tau$ the
conjugation of $\fg$ with respect to $\frak u(N)$. (Explicitly,
$\tau (\xi) = -\xi^{*}$ for $\xi\in \fg.$)  Clearly, the map
$$s = -\tau :\fg^{\Bbb{R}}\longrightarrow \fg^{\Bbb{R}}\eqno(5.2)$$
is an involutive Lie algebra anti-morphism satisfying $s(\fh) = \fh.$
In the following, the connected and simply-connected Lie groups
which integrate $\fg^{\Bbb{R}}$, $\fh^{\Bbb{R}}$ will be denoted
by $G^{\Bbb{R}}$ and $H^{\Bbb{R}}$ respestively.  Then the Lie
group anti-morphism 
$S:G^{\Bbb{R}}\longrightarrow G^{\Bbb{R}}$
corresponding to $s$ is
given by $S(g) = g^{*}$ for $g\in G^{\Bbb{R}}$.

In what follows, $U$ will denote  a fixed connected component of
$$\lbrace q=diag(q_1,\cdots q_N)\in \fh\mid \sinh\left({1\over 2}(q_{i}-q_{j})
 \right)
 \neq 0 \,\,\,\hbox{for all i and j}\rbrace.$$
We consider the solution $R:U\longrightarrow L(\fg,\fg)$ of the mDYBE, 
given by
$$R(q)\xi = -{1\over 2}\sum_{i\neq j} \coth\left({1\over 2}(q_i-q_j)\right)
  \xi_{ij} e_{ij},\eqno(5.3)$$
From this formula,
it is clear that $R$ induces a map $U\longrightarrow L(\fg^{\Bbb{R}},
\fg^{\Bbb{R}})$ which is a solution of the mDYBE for the pair 
$(\fg^{\Bbb{R}},\fh^{\Bbb{R}})$.  We shall denote this map also
by $R$ and from now onwards we shall only consider $R$ as
a map $U\longrightarrow L(\fg^{\Bbb{R}},
\fg^{\Bbb{R}})$.  We now equip the
trivial Lie groupoid
$$\Gamma = U\times G^{\Bbb{R}}\times U \eqno(5.4)$$
with the coboundary dynamical Poisson structure associated
to $R$.  Since $H^{\Bbb{R}}$ is abelian, its action on 
$\Gamma$ is given by
$${\Cal B}:H^{\Bbb{R}}\times \Gamma\longrightarrow \Gamma,\,\,
  {\Cal B}_{h}(u,g,v) = (u, hgh^{-1},v).\eqno(5.5)$$

\proclaim
{Proposition 5.1}  The map
$$\Sigma: (\Gamma, \{\, \cdot, \cdot \,\}_{R})\longrightarrow
  (\Gamma, \{\, \cdot, \cdot \,\}_{R}), (u,g,v)\mapsto
  (\bar v, g^{*}, \bar u)\eqno(5.6)$$
is a Poisson involution with stable locus
$$\Gamma^{\Sigma} = \lbrace (u,g,\bar u)\in \Gamma \mid g = g^{*} \rbrace 
.\eqno(5.7)$$
Hence $(\Gamma,\{\cdot,\cdot\}_{R},\Sigma)$ is a symmetric coboundary
dynamical Poisson groupoid.
\endproclaim

\demo
{Proof} Using the pairing $(\xi,\eta)_{\Bbb{R}} = 2\,Re\,tr(\xi\eta)$
on $\fg^{\Bbb{R}}$, it is straight forward to show that
$s^{*}_{\fh^{\Bbb{R}}}(u) = s_{\fh^{\Bbb{R}}}(u) = \bar u$
for $u\in U$.  By a similar calculation, we also have
$s^* = s$.  From this, it is easy to show that
$s\circ R(q)\circ s^{*} = -R(\bar q)$.  Hence it follows
from Proposition 3.12 that $\Sigma$ is a Poisson involution.
\pf
\enddemo

Let $T$ be the subgroup of $H^{\Bbb{R}}$ consisting of unitary
diagonal matrices and let $\frak t = Lie (T)$.  

\proclaim
{Proposition 5.2} (a) For all $d\in T$,
${\Cal B}_{d}\circ \Sigma = \Sigma \circ {\Cal B}_{d}.$
\smallskip
\noindent (b) $q-\bar q \in \frak t$\,\,\, for all $q\in U.$
\endproclaim 

\demo
{Proof} (a) From the definition of ${\Cal B}$ and $\Sigma$, we have
${\Cal B}_{d}\circ \Sigma = \Sigma \circ {\Cal B}_{d}$ for $d\in T$
iff $dg^{*}d^{*} = (dg^{*}d^{*})^{*}$.  But the latter is obvious.
\smallskip
\noindent (b) This assertion is clear.
\pf
\enddemo

Since we are dealing with the case in which the realization map is
the identity map on $\Gamma,$ it follows from Propositions 5.1 and
5.2 that assumptions G1-G4 in Section 3 are satisfied.  Hence we have
the following result by Proposition 3.19.

\proclaim
{Corollary 5.3} The stable locus $\Gamma^{\Sigma}$ is a Hamiltonian
$T$-space with equivariant momentum map
${\widetilde{\alpha}}-{\widetilde{\beta}} = \alpha - \beta\mid
\Gamma^{\Sigma}$.
\endproclaim

\definition
{Definition 5.4} The spin Ruijsenaars-Schneider models associated
to $R$ are the Hamiltonian systems on $\Gamma$ generated by
functions in $\bp^{*}_{2}I(G^{\Bbb R})$.  The symmetric space spin 
Ruijsenaars-Schneider models are the corresponding
Hamiltonian systems on $\Gamma^{\Sigma}$
generated by functions in $\bp^{*}_{2}I((G^{\Bbb R})^{S}).$
Here we have used the same symbol $\bp_{2}$ to denote the
projection map from $\Gamma$ to $G^{\Bbb{R}}$ and its
restriction from $\Gamma^{\Sigma}$ to $(G^{\Bbb{R}})^{S}.$
\enddefinition

Note that in the case under consideration, we have
$ {({\widetilde{\alpha}}-{\widetilde{\beta}})}^{-1}(0)\simeq 
U_{s}\times (G^{\Bbb R})^{S},$ where
$U_{s}$ consists of real diagonal matrices in $U$ and
$(G^{\Bbb R})^{S}$ consists of Hermitian matrices in $G^{\Bbb{R}}.$
As we are identifying $(\fg^{\Bbb R})^*$ with $\fg^{\Bbb R}$ using
the pairing $(\cdot,\cdot)_{\Bbb R}$, it follows from (3.7) that
the Poisson structure on $\Gamma^{\Sigma}$ is given by
$$\eqalign {\{\varphit,\psit\}_{\Gamma^{\Sigma}} (u,g,\bar u)
=& - 2(\iota\delta_1 \varphit,\, D\psit)_{\Bbb R}
 +2 (\iota\delta_1 \psit,\, D\varphit)_{\Bbb R}
  -2(R(u) D\varphit, D\psit)_{\Bbb R}\cr}\eqno(5.8)$$
where 
$$\delta_{1}\varphit:= {1\over 2}(\delta_{1}\varphi + 
     \overline{\delta_{2}\varphi}),
        \quad D\varphit:={1\over 2}(D\varphi + (D'\varphi)^{*}).\eqno(5.9)$$
Hence we obtain the following result by applying Proposition 3.13.

\proclaim
{Proposition 5.5} Let $f(g) = 2 Re\, tr (g)$, $g\in G^{\Bbb R}$ and
let $F = Pr_{2}^{*}f$.  Then the restriction of the Hamiltonian 
equations of motion generated by $\widetilde F$ to the invariant
manifold $U_{s}\times (G^{\Bbb R})^{S}$ are given by
$$\aligned
         & \dot q = \Pi_{\fh^{\Bbb R}}\,\, g, \\
         & \dot g = g (R(q) g) - (R(q) g) g.
\endaligned
$$
In terms of the components of $q$ and $g$, these read:
$$\ddot q_{i} =  \dot g_{ii}
  = {1\over 2}\sum_{k\neq i} \coth ((q_{i}-q_{k})/2) g_{ik} g_{ki},
  \eqno(5.10a)$$
$$\eqalign{\dot g_{ij} =& {1 \over 2} \coth ((q_{i}-q_{j})/2) g_{ij}
    (g_{jj}-g_{ii})\cr
    & + {1\over 2} \sum_{k\neq i,j} \bigl( \coth((q_{i}-q_{k})/2)
    - \coth((q_{k}-q_{j})/2)\bigr) g_{ik} g_{kj}, \quad i\neq j\cr}
   \eqno(5.10b)
$$
\endproclaim
Note that the  equations in (5.10) for some special choice of
$g$ are exactly the ones
derived by Braden and Hone in \c{BH} from the $N$-soliton solutions
of the $A^{(1)}_n$ affine Toda field theory with purely imaginary
coupling constant (cf.\c{KZ}).  For the convenience of the reader, let us 
recall that the equations of motion of the $A^{(1)}_n$ affine
Toda field theory (with imaginary coupling constant $i\beta$) are given by
$$\partial_{+}\partial_{-}\phi_{j} + {m^{2}\over {2i\beta}}\left
  (e^{i\beta (\phi_{j}-\phi_{j+1})} -e^{i\beta (\phi_{j-1}-\phi_{j})}\right)
  = 0, \quad j=0,\ldots, n. \eqno(5.11)$$
Here, 
$\partial_{\pm}$ denotes differentiation with respect to the light-cone
coordinates $x_{\pm} = {1\over \sqrt 2}(t\pm x)$ and the indices
$j$ are taken modulo $n+1$.  In \c{BH}, the authors were dealing
with the solitonic sector of the theory, so they assumed in addition
that $\sum_{j=0}^{n} \phi_{j} = 0$. Starting from the $N$-soliton
solutions of (5.10) as derived by Hollowood \c{H}:
$$e^{i\beta \phi_{j}} = \frac {\tau_{j+1}}{\tau_{j}},\eqno(5.12)$$ 
where $\tau_{j}$ are the tau functions, Braden and Hone began
by rewriting $\tau_{j}$ in determinantal form.  As it turned out,
they found that
$$\tau_{j} = det \left( 1 + e^{ij\Theta/ 2}Ve^{ij\Theta/ 2}\right)
  \eqno(5.13)$$
where $V$ is an invertible skew-Hermitian matrix depending on
$x_{\pm}$ (and the $2N$ soliton parameters) satisfying 
$$\dot V = {1\over 2}(\Lambda V + V \Lambda),\quad \cdot  =\partial_{\pm}
  \eqno(5.14)$$
In (5.13) and (5.14), $\Theta = diag (\theta_{1},\ldots, \theta_{N})$
and
$$\Lambda = diag \left(\pm \sqrt{2} m \,\exp(\mp \eta_{1}) sin (\theta_{1}/2),
  \ldots, \pm \sqrt{2} m \,\exp(\mp \eta_{N}) sin (\theta_{N}/2) \right)
  \eqno(5.15)$$
where $\theta_{j}$ are discrete parameters associated with
the solitons taking
values in $\lbrace {2\pi k\over n+1}\mid k = 1,\ldots, n\rbrace$,
and $\eta_{j}$ are the rapidities.  As $V$ is 
skew-Hermitian and invertible, there exists a unitary matrix $U$ 
(unique up to transformations $U\rightarrow \delta U$, where $\delta\in T$)
which diagonalizes $V$, i.e.
$$ie^{q} = UVU^{*}\eqno(5.16)$$
where $q$ is real diagonal.
Using $U$, define an invertible Hermitian matrix $g$ by:
$$g = U\Lambda U^{*}.\eqno(5.17)$$
Then under the evolution as defined by (5.14), the authors in
\c{BH} showed that $q$ and $g$ satisfy (5.10)!  Of course,
in this context, the variable $g$ depends on the choice of $U$.
Clearly, the system which is independent of such a choice is the
corresponding reduced system on the Poisson quotient
$U_{s}\times ((G^{\Bbb R})^{S}/T).$  As a consequence of 
Theorem 3.20, we therefore conclude that the reduced system
has $N$ Poisson commuting integrals.
\smallskip
\noindent{\bf Remark 5.7.} There are several questions which
we have not addressed in our discussion above.  One of these
has to do with the nature of the transformation between
the $2N$ soliton parameters and the dynamical variables
in $U_{s}\times ((G^{\Bbb R})^{S}/T).$  On the other hand, there
should be corresponding results for the soliton solutions
associated with the other affine Lie algebras in \c{OTU}.
Of course, there is also the question of Liouville integrability.
We hope to return to these questions in future work.

\enddocument